\documentclass[a4paper]{article}
\usepackage[utf8x]{inputenc}
\usepackage[left=3cm,right=3cm,top=3cm,bottom=3cm]{geometry}
\usepackage[linktoc=all,hidelinks]{hyperref}
\usepackage{cite}

\usepackage{amsmath}
\usepackage{amsfonts}
\usepackage{amssymb}

\usepackage{tensor}
\usepackage{youngtab}
\usepackage{slashed}

\newcommand{\dx}[1][D]{\! d^{#1}\! x}
\newcommand{\dtdx}[1][d]{\! dt \, d^{#1}\! x}

\newcommand{\col}[2]{\begin{pmatrix} #1 \vspace{5pt} \\ #2 \end{pmatrix}}

\newcommand{\cB}{\mathcal{B}}
\newcommand{\cG}{\mathcal{G}}
\newcommand{\cH}{\mathcal{H}}

\newcommand{\pd}{\partial}
\newcommand{\lap}{\,\Delta}

\newcommand{\hc}{\text{h.c.}}

\newcommand{\nn}{\nonumber}

\numberwithin{equation}{section}

\begin{document}

\thispagestyle{empty}
\begin{center}

\vspace*{50pt}
{\LARGE \bf Prepotentials for linearized supergravity}

\vspace{30pt}
{Victor Lekeu${}^{\, a}$ and Amaury Leonard${}^{\, a,b}$}

\vspace{10pt}
\texttt{vlekeu@ulb.ac.be, amleonar@ulb.ac.be}

\vspace{20pt}
\begin{enumerate}
\item[${}^a$] {\sl \small
Universit\'e Libre de Bruxelles and International Solvay Institutes,\\
ULB-Campus Plaine CP231, B-1050 Brussels, Belgium}
\item[${}^b$] {\sl \small
Max-Planck-Institut f\"{u}r Gravitationsphysik (Albert-Einstein-Institut),\\
Am M\"{u}hlenberg 1, DE-14476 Potsdam, Germany}
\end{enumerate}

\vspace{50pt}
{\bf Abstract} 
\end{center}

\noindent
Linearized supergravity in arbitrary dimension is reformulated into a first order formalism which treats the graviton and its dual on the same footing at the level of the action. This generalizes previous work by other authors in two directions: 1) we work in arbitrary space-time dimension, and 2) the gravitino field and supersymmetry are also considered. This requires the construction of conformally invariant curvatures (the Cotton fields) for a family of mixed symmetry tensors and tensor-spinors, whose properties we prove (invariance; completeness; conformal Poincaré lemma). We use these geometric tools to solve the Hamiltonian constraints appearing in the first order formalism of the graviton and gravitino: the constraints are solved through the introduction of prepotentials enjoying (linearized) conformal invariance. These new variables (two tensor fields for the graviton, one tensor-spinor for the gravitino) are injected into the action and equations of motion, which take a geometrically simple form in terms of the Cotton tensor(-spinors) of the prepotentials. In particular, the equations of motion of the graviton are equivalent to twisted self-duality conditions. We express the supersymmetric transformations of the graviton and gravitino into each other in terms of the prepotentials. We also reproduce the dimensional reduction of supergravity within the prepotential formalism. Finally, our formulas in dimension five are recovered from the dimensional reduction of the already known prepotential formulation of the six-dimensional $\mathcal{N}=(4,0)$ maximally supersymmetric theory.

\newpage

\setcounter{tocdepth}{2}
\tableofcontents

\newpage

\section{Introduction}

Electric-magnetic duality has been the object of growing interest, as it appears in a large variety of theories. Initially observed at the level of the equations of motion and Bianchi identities, the field and its dual can be made to appear on an equal footing at the level of the action \cite{Deser:1976iy,Deser:1981fr,Henneaux:1988gg,Schwarz:1993vs,Deser:1997mz,Deser:1997se,Hillmann:2009zf,Bunster:2011aw,Henneaux:2017kbx} by going to the first order formalism in terms of unconstrained Hamiltonian variables. This rewriting of the action entails a loss of explicit space-time covariance, which could be complementary to duality \cite{Bunster:2012hm}. The action is nevertheless covariant.

A similar duality arises for linearized gravity \cite{Hull:2000zn,Hull:2001iu,Hull:2000rr,West:2001as,Nieto:1999pn,Bekaert:2002jn,Boulanger:2003vs}. In space-time dimension $D$, the ``dual graviton" is a tensor of mixed Young symmetry type $(D-3,1)$ (see \cite{Curtright:1980yk} for early work on mixed symmetry gauge fields).
Gravitational duality invariance is involved in the infinite-dimensional Kac-Moody algebras $E_{11}$ \cite{West:2001as} or $E_{10}$ \cite{Damour:2002cu} which are conjectured to be the fundamental symmetries of supergravity and (for $E_{10}$) the zero tension limit of string theory \cite{Gross:1988ue}. Understanding the role played by the dual of the graviton could be a step toward unravelling of these ``hidden" infinite-dimensional symmetries.
It can also be understood geometrically from the dimensional reduction on a torus of the exotic six-dimensional models of \cite{Hull:2000zn,Hull:2000rr}. Generally speaking, a formalism which treats the fields and their dual on the same footing is the rewriting of the equations of motion as twisted self-duality conditions \cite{Cremmer:1979up,Cremmer:1997ct,Cremmer:1998px,Bunster:2011qp} which, in the case of electromagnetism, equate the electric field of the potential vector to the magnetic field of its dual, and reciprocally (up to a sign)\footnote{It is to be observed that in dimension greater than four the field and its dual no longer share the same symmetry type, so that the invariance under $SO(2)$ duality rotations is absent in higher dimensions, where only the twisted self-duality subsists.}. \\

As in the case of electromagnetism, duality invariance can be realized at the level of the action by going to the first order formalism and working with unconstrained Hamiltonian variables: these are the prepotentials, which appear in pair.
Linearized gravity has been brought to this form in dimension four and five in  \cite{Henneaux:2004jw,Bunster:2012km,Bunster:2013oaa}. In this paper, we generalize this reformulation to arbitrary dimension and give it a systematic geometric base.

As was observed before, the prepotentials are always found to enjoy a larger gauge invariance than the corresponding covariant fields used in second order formalism: their gauge transformations include both linearized diffeomorphisms and local Weyl rescalings. This observation leads us to investigate the associated conformal geometry and the construction of a complete set of invariant conformal curvatures, which are the Cotton tensors, whose properties we establish (invariance, completeness). The Hamiltonian analysis and the rewriting of the equations of motion as twisted self-duality conditions are greatly clarified by the use of these conformal techniques. Our treatment gives a formulation of linearized supergravity that treats the graviton and its dual on an equal footing in arbitrary dimension. In particular, this can be applied to eleven-dimensional supergravity, where it could help exhibiting the infinite-dimensional symmetry this theory is speculated to enjoy.

We also apply this treatment to the gravitino, for which the appropriate conformal geometry is built in arbitrary dimension. This allows us to write its first order action and equations of motion in terms of a prepotential enjoying Weyl invariance - as usual, at the cost of explicit space-time covariance. Finally, we express the supersymmetry transformations of the graviton and the gravitino in linearized supergravity using these prepotential variables, generalizing the work of \cite{Bunster:2012jp}\footnote{A similar reformulation was already carried out for the four-dimensional linearized ``hypergravity" (a supermultiplet combining the graviton and a spin $5/2$ field), without systematic use of conformal tools, in \cite{Bunster:2014fca}.}.

As far as supergravity is concerned, a covariant action principle is available for the nonlinear theory. However, this is not the case for the intriguing $\mathcal{N} = (4,0)$ maximally supersymmetric theory in six dimensions \cite{Hull:2000zn,Hull:2000rr}, whose six-dimensional action was only recently obtained and involves prepotentials \cite{Henneaux:2016opm,Henneaux:2017xsb}. We show that it reduces in five dimensions to the actions obtained in this paper for linearized supergravity.\\

This paper is organized as follows. We begin, in section \ref{sec:conformal}, by developing our conformal tools. We consider tensor and tensor-spinor fields of specific types of mixed symmetry which are endowed with an invariance under both generalized gauge transformations (linearized diffeomorphisms) and local Weyl transformations, and build a complete set of curvatures invariant under these transformations.

This geometric apparatus is then used to carry out the Hamiltonian analysis of linearized supergravity in arbitrary dimension: section \ref{sec:graviton} treats the bosonic field and section \ref{sec:gravitino} the fermionic field. In both cases, the Hamiltonian constraints are obtained from the space-time break-up of the covariant action (enjoying a diffeomorphism invariance), and their resolution is shown to be expressed with the fields whose conformal geometry was analysed in section \ref{sec:conformal}: these are the prepotentials, indeed found to enjoy a gauge invariance combining diffeomorphism and Weyl transformations. Two prepotentials naturally appear in the bosonic case, associated to the field and its conjugate momentum, and only one in the fermionic case.

We then turn, in section \ref{sec:susy}, to the supersymmetric invariance. We begin with the covariant form of the supersymmetry transformations and rewrite them in terms of the prepotentials.

The appendices contain a series of additional details. Appendix \ref{app:cotton} gathers the proofs of the key properties of the Cotton tensor(-spinor). In appendix \ref{app:lowspin}, we review the analogous treatment of the scalar, Dirac and vector fields. This is used in appendix \ref{app:dimred}, in which we perform the Kaluza-Klein dimensional reduction (from $D+1$ to $D$ dimensions) of linearized supergravity in our formalism, reproducing the well-known results. Appendix \ref{app:40} derives the dimensional reduction of six-dimensional self-dual fields appearing in the $\mathcal{N} = (4,0)$ maximally supersymmetric theory of \cite{Hull:2000zn,Hull:2000rr}. Finally, appendix \ref{app:gammamatrices} contains conventions and useful identities involving gamma matrices in arbitrary dimension.

\section{Conformal geometry}
\label{sec:conformal}

The goal of this section is to analyse the properties of certain conformal fields in $d$ Euclidean dimensions. In addition to their gauge transformations, those fields also enjoy local Weyl symmetries (hence the name ``conformal" by abuse of teminology). The geometric tensors for those symmetries are systematically constructed.

The cases we analyse here are those relevant for the prepotentials of the graviton and gravitino in $D = d+1$ spacetime dimensions. The pattern is by now well established and follows references  \cite{Bunster:2012km,Bunster:2013oaa,Henneaux:2015cda,Henneaux:2016zlu,Henneaux:2016opm,Henneaux:2017xsb}. The main feature of these cases is that the analogue of the Weyl tensor (the traceless part of the curvature tensors) identically vanishes: one must therefore construct the analogue of the Cotton tensor of three-dimensional gravity. These generalized Cotton tensors satisfy two important properties:
\begin{enumerate}
\item They completely control Weyl invariance: the Cotton tensor is zero if and only if the field can be written as a the sum of a gauge and a Weyl transformation.
\item They are divergenceless and also obey some appropriate trace condition. Conversely, any traceless, divergenceless tensor with the symmetries of the Cotton tensor can be written as the Cotton tensor of some field. (The ambiguities in determining this field are precisely the gauge and Weyl transformations that we consider.)
\end{enumerate}
The first of these properties is equivalent to the fact that any function of the fields that is invariant under gauge and Weyl transformations can be written as a function of the Cotton tensor only. The second property is the one that allows us to solve the constraints appearing in the Hamiltonian formulation of linearized supergravity. The proofs of those two properties are presented in appendix \ref{app:cotton}. They strongly rely on the generalized Poincaré lemmas of \cite{Olver_hyper,DuboisViolette:1999rd,DuboisViolette:2001jk,Bekaert:2002dt} for tensors of mixed Young symmetry.

\subsection{Bosonic $(d-2,\,1)$-field}
\label{app:confphi}

We first consider a bosonic field $\phi_{i_1 \dots i_{d-2} j }$ with the symmetries of the two-column $(d-2, \,1)$ Young tableau\footnote{In our conventions, the integers $(p,q)$ indicate the heights of the columns.}, i.e.
\begin{equation}
\phi_{i_1 \dots i_{d-2} j } = \phi_{[i_1 \dots i_{d-2}] j }, \quad \phi_{[i_1 \dots i_{d-2} j] } = 0 ,
\end{equation}
with the local gauge and Weyl symmetries
\begin{equation}\label{eq:phigaugeapp}
\delta \phi\indices{^{i_1 \dots i_{d-2}}_{j} } = \partial^{[i_1} M\indices{^{i_2 \dots i_{d-2}]}_j} + \partial_j A^{i_1 \dots i_{d-2}} + \partial^{[i_1} A\indices{_j^{i_2 \dots i_{d-2}]}} + \delta^{[i_1}_j B^{i_2 \dots i_{d-2}]},
\end{equation}
where $M$ has the $(d-3,1)$ mixed symmetry and $A$, $B$ are totally antisymmetric. The tensors $M$ and $A$ are the usual gauge parameters for a mixed symmetry field while $B$ is the parameter for the local Weyl transformations of $\phi$.

\paragraph{Einstein tensor.} The Einstein tensor is obtained by taking a curl of $\phi$ on both groups of indices,
\begin{equation}
G\indices{^{i_1 \dots i_{d-2}}_{j}} [\phi] = \partial^k \partial_m \phi\indices{^{l_1 \dots l_{d-2}}_{n} } \varepsilon^{mni_1 \dots i_{d-2}} \varepsilon_{jkl_1 \dots l_{d-2}} .
\end{equation}
It is invariant under the gauge symmetries parametrized by $M$ and $A$. It is identically divergenceless (on both groups of indices) and also has the $(d-2,1)$ Young symmetry. The converse of these properties is also true and is easily proven using the generalized Poincaré lemma of \cite{Bekaert:2002dt} for mixed symmetry fields:
\begin{itemize}
\item The condition $G[\phi] = 0$ implies that the field is is pure gauge,
\begin{equation}\label{eq:gaugeEinsteinphi}
G\indices{^{i_1 \dots i_{d-2}}_{j}} [\phi] = 0 \quad\Leftrightarrow\quad \phi\indices{^{i_1 \dots i_{d-2}}_{j} } = \partial^{[i_1} M\indices{^{i_2 \dots i_{d-2}]}_j} + \partial_j A^{i_1 \dots i_{d-2}} + \partial^{[i_1} A\indices{_j^{i_2 \dots i_{d-2}]}}
\end{equation}
for some $A$ and $M$ with the appropriate symmetry.
\item Any divergenceless tensor of $(d-2,1)$ symmetry is the Einstein tensor of some $(d-2,1)$ field,
\begin{equation}
\partial_{i_1} T\indices{^{i_1 \dots i_{d-2}}_{j}} = 0 \quad \Leftrightarrow \quad T = G[\phi] \;\text{ for some } \phi.
\end{equation}
(The divergencelessness of $T$ in the $j$ index follows from the cyclic identity $T_{[i_1 \dots i_{d-2} j]} = 0$.)
\end{itemize}

\paragraph{Schouten tensor.}  The Einstein tensor is not invariant under the Weyl symmetries parametrized by $B$, under which it transforms as
\begin{equation}\label{eq:varGphi}
\delta G\indices{^{i_1 \dots i_{d-2}}_{j}} [\phi] = \partial^k \partial_m B^{l_2 \dots l_{d-2} } \varepsilon^{mni_1 \dots i_{d-2}} \varepsilon_{jknl_2 \dots l_{d-2}} = - (d-1)! \partial^k \partial_m B^{l_2 \dots l_{d-2} } \delta^{m i_1 \dots i_{d-2}}_{jkl_2 \dots l_{d-2}} .
\end{equation}
For the trace $G^{i_2 \dots i_{d-2}} [\phi] \equiv G\indices{^{j i_2 \dots i_{d-2}}_{j}} [\phi]$, this implies
\begin{equation}\label{eq:vartraceGphi}
\delta G^{i_2 \dots i_{d-2}} [\phi] = \partial^k \partial_m B^{l_2 \dots l_{d-2} } \varepsilon^{mnj i_2 \dots i_{d-2}} \varepsilon_{jknl_2 \dots l_{d-2}} = 2(d-2)! \partial^k \partial_m B^{l_2 \dots l_{d-2} } \delta^{m i_2 \dots i_{d-2}}_{k l_2 \dots l_{d-2}}
\end{equation}
From the Einstein tensor and its trace, one can define the Schouten tensor
\begin{equation}\label{eq:SofGphi}
S\indices{^{i_1 \dots i_{d-2}}_{j}} [\phi] = G\indices{^{i_1 \dots i_{d-2}}_{j}} [\phi] - \frac{d-2}{2} \,\delta^{[i_1}_j G^{i_2 \dots i_{d-2}]}[\phi]
\end{equation}
which has the important property of transforming simply as
\begin{equation}\label{eq:transfSchphi}
\delta S\indices{^{i_1 \dots i_{d-2}}_{j}} [\phi] = - (d-2)!\,\partial_j\partial^{[i_1} B^{i_2 \dots i_{d-2}]}
\end{equation}
under a Weyl transformation, as follows from \eqref{eq:varGphi} and \eqref{eq:vartraceGphi}.

The relation between the Einstein and the Schouten can be inverted to
\begin{equation}
G\indices{^{i_1 \dots i_{d-2}}_{j}} [\phi] = S\indices{^{i_1 \dots i_{d-2}}_{j}} [\phi] - (d-2) \,\delta^{[i_1}_j S^{i_2 \dots i_{d-2}]}[\phi],
\end{equation}
where $S^{i_2 \dots i_{d-2}} [\phi] \equiv S\indices{^{j i_2 \dots i_{d-2}}_{j}}$ is the trace of the Schouten tensor.
The divergencelessness of $G$ is then equivalent to
\begin{equation}\label{eq:bianchiSphi}
\partial_{i_1} S\indices{^{i_1 \dots i_{d-2}}_{j}} [\phi] - \partial_j S^{i_2 \dots i_{d-2}} = 0 .
\end{equation}
It also implies the identity
\begin{equation}
\partial^j S\indices{^{i_1 \dots i_{d-2}}_{j}} [\phi] - (d-2) \partial^{[i_1} S^{i_2 \dots i_{d-2}]} = 0
\end{equation}
because $G$ is divergenceless on its last index.

\paragraph{Cotton tensor.} The Cotton tensor is defined as
\begin{equation}\label{eq:cottonphi}
D^{i_1 \dots i_{d-2} \, j_1 \dots j_{d-2}}[\phi] = \varepsilon^{i_1 \dots i_{d-2}kl} \partial_k S\indices{^{j_1 \dots j_{d-2}}_{l}} [\phi] .
\end{equation}
Because of the transformation law \eqref{eq:transfSchphi} of the Schouten tensor, it is invariant under the full gauge and Weyl transformations \eqref{eq:phigaugeapp}. Moreover, $D[\phi] = 0$ implies that the field takes the form \eqref{eq:phigaugeapp} (see appendix \ref{app:cotton}). It also satisfies the following properties:
\begin{itemize}
\item It has the $(d-2,d-2)$ Young symmetry,
\begin{equation}\label{eq:cyclicDphi}
D_{[i_1 \dots i_{d-2} \, j_1] j_2 \dots j_{d-2}}[\phi] = 0,
\end{equation}
\item It is divergenceless,
\begin{equation}
\partial^{i_1}D_{i_1 \dots i_{d-2}\, j_1 \dots j_{d-2}}[\phi] = 0,
\end{equation}
\item Its complete trace vanishes,
\begin{equation}
D\indices{^{i_1 \dots i_{d-2}}_{i_1 \dots i_{d-2}}}[\phi] = 0 .
\end{equation}
\end{itemize}
The first of these properties is equivalent to the identity \eqref{eq:bianchiSphi}. The second is evident from the definition of $D$; divergencelessness on the second group of indices then follows from the cyclic identity \eqref{eq:cyclicDphi}. The last property follows from the cyclic identity for the Schouten tensor.
Conversely, any tensor that satisfies these three properties must be the Cotton tensor of some $(d-2,1)$ field.

Note that the transformation property \eqref{eq:transfSchphi} of the Schouten implies that the tensor
\begin{equation}
D'_{ij}[\phi] = \varepsilon_{ikl_1 \dots l_{d-2}} \partial^k S\indices{^{l_1 \dots l_{d-2}}_j}[\phi]
\end{equation}
is also invariant under Weyl transformations of $\phi$. (With respect to definition \eqref{eq:cottonphi}, $D'$ is obtained by taking the curl of $S$ on the other group of indices.) Therefore, it must be a function of the Cotton tensor. Indeed, using the cyclic identity $S_{[i_1 \dots i_{d-2} | j]} = 0$, one finds
\begin{equation}\label{eq:dprimephi}
D\indices{^{\prime i}_j}[\phi] = (d-2) D\indices{^{ik_2 \dots k_{d-2}}_{jk_2 \dots k_{d-2}}}[\phi] .
\end{equation}

\subsection{Bosonic $(d-2,\,d-2)$-field}
\label{app:confP}

We consider now a bosonic field $P_{i_1 \dots i_{d-2} \, j_1 \dots j_{d-2} }$ with the symmetries of the $(d-2, \,d-2)$ Young tableau, i.e.
\begin{equation}
P_{i_1 \dots i_{d-2} \, j_1 \dots j_{d-2} } = P_{[i_1 \dots i_{d-2}] \, [j_1 \dots j_{d-2}] }, \quad P_{[i_1 \dots i_{d-2} \, j_1] j_2 \dots j_{d-2} } = 0 .
\end{equation}
This field has the gauge symmetries
\begin{align} \label{eq:Pgaugeapp}
\delta P\indices{^{i_1 \dots i_{d-2}}_{j_1 \dots j_{d-2} }} = \; &\alpha\indices{^{i_1 \dots i_{d-2}}_{[j_1 \dots j_{d-3}, j_{d-2}] }} + \alpha\indices{_{j_1 \dots j_{d-2}}^{[i_1 \dots i_{d-3}, i_{d-2}] }} \nonumber \\
&+ \delta^{i_1 \dots i_{d-2}}_{j_1 \dots j_{d-2}} \, \xi ,
\end{align}
where $\alpha$ has the $(d-2,\,d-3)$ Young symmetry and the comma denotes the derivative. The $\alpha$ transformations are the usual gauge tranformations for a $(d-2, \,d-2)$ mixed symmetry field, while the $\xi$ transformations are the local Weyl symmetries in this case.

\paragraph{Einstein tensor.} The Einstein tensor of $P$ is defined as
\begin{equation}
G_{ij}[P] = \varepsilon_{ikm_1 \dots m_{d-2}} \varepsilon_{jln_1 \dots n_{d-2}} \partial^k \partial^l P^{m_1 \dots m_{d-2} \, n_1 \dots n_{d-2} } .
\end{equation}
It is invariant under the $\alpha$ gauge symmetries. It is also symmetric and divergenceless. Conversely, the condition $G_{ij}[P] = 0$ implies that $P$ is pure gauge, and any symmetric divergenceless tensor is the Einstein tensor of some field $P$ with the $(d-2,d-2)$ symmetry. These two theorems are easily proven using the Poincaré lemma of \cite{DuboisViolette:1999rd,DuboisViolette:2001jk} for rectangular Young tableaux.

\paragraph{Schouten tensor.} Under a Weyl transformation, we have
\begin{equation}
\delta G_{ij}[P] = (d-2)! (\delta_{ij} \lap \xi - \partial_i \partial_j \xi )
\end{equation}
and $\delta G[P] = (d-1)! \lap \xi$ for the trace $G[P] \equiv G\indices{^i_i}[P]$. The Schouten tensor is then defined as
\begin{equation}
S_{ij}[P] = G_{ij}[P] -  \frac{1}{d-1} \, \delta_{ij}\, G[P] .
\end{equation}
It transforms in a simple way under Weyl transformations,
\begin{equation}\label{eq:deltaSP}
\delta S_{ij}[P] = -(d-2)! \,\partial_i \partial_j \xi .
\end{equation}
It is also symmetric. One can invert the relation between the Schouten and the Einstein as $G_{ij} = S_{ij} - \delta_{ij} S$. Therefore, the Schouten tensor satisfies
\begin{equation}\label{eq:SPdiv}
\partial^i S_{ij} = \partial_j S
\end{equation}
because $G_{ij}$ is divergenceless.

\paragraph{Cotton tensor.} The Cotton tensor is then defined as
\begin{equation}
D_{i_1 \dots i_{d-2} \, j }[P] = \varepsilon_{i_1 \dots i_{d-2} kl } \partial^k S\indices{^l_j}[P] .
\end{equation}
It is invariant under all the gauge symmetries \eqref{eq:Pgaugeapp} as a consequence of \eqref{eq:deltaSP}, and the condition $D[P] = 0$ implies that $P$ is of the form \eqref{eq:Pgaugeapp}.
It is a tensor of $(d-2,1)$ mixed symmetry,
\begin{equation}\label{eq:DPsym}
D_{i_1 \dots i_{d-2} \, j } = D_{[i_1 \dots i_{d-2}] \, j }, \quad D_{[i_1 \dots i_{d-2} \, j]} = 0.
\end{equation}
The second of these equations is equivalent to \eqref{eq:SPdiv}. Moreover, because the Schouten is symmmetric, the Cotton is traceless,
\begin{equation}
D\indices{^j_{i_2 \dots i_{d-2} \, j }} = 0 .
\end{equation}
Lastly, the Cotton is also identically divergenceless on the first group of indices,
\begin{equation}
\partial^{i_1} D_{i_1 \dots i_{d-2} \, j } = 0 .
\end{equation}
The divergencelessness on the last index, $\partial^j D_{i_1 \dots i_{d-2} \, j } = 0$, is then a consequence of this and the second of \eqref{eq:DPsym}. Conversely, any $(d-2,1)$ field that is traceless and divergenceless is the Cotton tensor of some $P$.

\subsection{Fermionic $(d-2)$-form}
\label{app:conffermion}

We consider now an antisymmetric tensor-spinor $\chi_{i_1 \dots i_{d-2}}$, with the gauge and Weyl transformations
\begin{equation} \label{eq:chigaugeapp}
\delta \chi_{i_1 \dots i_{d-2}} = (d-2) \partial_{[i_1} \eta_{i_2 \dots i_{d-2}]} + \gamma_{i_1 \dots i_{d-2}} \rho
\end{equation}
where $\eta_{i_1 \dots i_{d-3}}$ and $\rho$ are spinor fields (see appendix \ref{app:gammamatrices} for our gamma matrix conventions).

\paragraph{Einstein tensor.} The invariant tensor for the $\eta$ transformations is of course the curl
\begin{equation}
G_i[\chi] = \varepsilon_{ijk_1\dots k_{d-2}} \partial^j \chi^{k_1 \dots k_{d-2}} ,
\end{equation}
which we also call ``Einstein tensor" by analogy with the previous cases. It satisfies the following properties, which are easily proved using the usual Poincaré lemma with a spectator spinor index:
\begin{itemize}
\item It is invariant under the $\eta$ gauge transformations. Conversely, $G_i[\chi] = 0$ implies that $\chi_{i_1 \dots i_{d-2}}$ is pure gauge,
\begin{equation}
G_i[\chi] = 0 \quad\Leftrightarrow\quad \chi_{i_1 \dots i_{d-2}} = (d-2) \partial_{[i_1} \eta_{i_2 \dots i_{d-2}]} \;\text{ for some } \eta.
\end{equation}
\item It is identically divergenceless, $\partial^i G_i = 0$. Conversely, any divergenceless vector-spinor is the Einstein tensor of some antisymmetric tensor-spinor $\chi_{i_1 \dots i_{d-2}}$,
\begin{equation}
\partial^i T_i = 0 \quad \Leftrightarrow \quad T_i = G_i[\chi] \;\text{ for some } \chi.
\end{equation}
\end{itemize}

\paragraph{Schouten tensor.} The Einstein tensor is not invariant under Weyl transformations. We have
\begin{equation}
\delta G^i[\chi] = \varepsilon^{ijl_1 \dots l_{d-2}} \gamma_{l_1\dots l_2}\partial_j \rho  = i^{m+1} (d-2)!\, \gamma^{ij} \gamma_0 \hat{\gamma}\, \partial_j \rho
\end{equation}
where $m = \lfloor (d+1)/2 \rfloor$ and the dimension-dependent matrix $\hat{\gamma}$ is defined in appendix \ref{app:gammamatrices}.

The Schouten tensor is then defined as
\begin{equation}
S_i[\chi] = \frac{1}{d-1} \left( \gamma_{ij} - (d-2) \delta_{ij} \right) G^j[\chi] .
\end{equation}
Using the gamma matrix identity \eqref{eq:gammadelta}, one can see that it transforms simply as a total derivative under Weyl rescalings,
\begin{equation}
\delta S_i [\chi] = \partial_i \nu, \qquad \nu = i^{m+1} (d-2)! \gamma_0 \hat{\gamma}\, \rho .
\end{equation}
The Einstein can be written in terms of the Schouten as
\begin{equation}\label{eq:GofSchi}
G_i[\chi] = \gamma_{ij} S^j[\chi] ,
\end{equation}
which implies that the Schouten identically satisfies
\begin{equation}\label{eq:dSfermionic}
\gamma_{ij} \partial^i S^j[\chi] = 0 .
\end{equation}

\paragraph{Cotton tensor.} The invariant tensor for Weyl transformations is the Cotton tensor
\begin{equation}\label{eq:defcottonchi}
D_{i_1 \dots i_{d-2}}[\chi] = \varepsilon_{i_1\dots i_{d-2}jk} \partial^j S^k[\chi] .
\end{equation}
It is identically divergenceless. Its complete gamma-trace also vanishes,
\begin{equation}\label{eq:cottongammatrace}
\gamma^{i_1 \dots i_{d-2}} D_{i_1 \dots i_{d-2}} = 0,
\end{equation}
as follows from \eqref{eq:dSfermionic} and identity \eqref{eq:gij}. Conversely, $D[\chi] = 0$ implies that $\chi$ takes the form \eqref{eq:chigaugeapp}, and any divergenceless, rank $d-2$ antisymmetric tensor-spinor satisfying the complete gamma-tracelessness condition \eqref{eq:cottongammatrace} is the Cotton tensor of some $\chi$. The proof of those properties is done in appendix \ref{app:cotton}.

\section{Graviton}
\label{sec:graviton}

The results of the previous sections allow us to solve the constraints appearing in the Hamiltonian formulation of linearized gravity. In doing so, the dynamical variables are expressed in terms of two fields $\phi_{i_1 \dots i_{d-2} j}$ and $P_{i_1 \dots i_{d-2} j_1 \dots j_{d-2}}$ of respective $(d-2,1)$ and $(d-2,d-2)$ Young symmetry, called ``prepotentials". This generalizes to arbitrary dimension the work of \cite{Henneaux:2004jw,Bunster:2012km} and \cite{Bunster:2013oaa}, to which our results reduce in $D = 4$ and $5$ respectively. A further improvement over that previous work is the complete rewriting of the action in terms of the relevant Cotton tensors, which makes its gauge and Weyl invariance manifest\footnote{References \cite{Bunster:2012km,Bunster:2013oaa} already contain this rewriting of the kinetic term in four and five dimensions, but not the rewriting of the Hamiltonian.}.

\subsection{Hamiltonian action}

The Pauli-Fierz Lagrangian for linearized gravity is
\begin{equation} \label{LPF}
\mathcal{L}_\text{PF} = -\frac{1}{4} \partial_\mu h_{\nu\rho} \partial^\mu h^{\nu\rho} + \frac{1}{2} \partial_\mu h^{\mu\nu} \partial^\rho h_{\rho\nu} - \frac{1}{2} \partial_\mu h^{\mu\nu} \partial_\nu h\indices{^\rho_\rho} + \frac{1}{4} \partial_\mu h\indices{^\nu_\nu} \partial^\mu h\indices{^\rho_\rho} .
\end{equation}
It can also be written as
\begin{equation}
\mathcal{L}_\text{PF} = - \frac{3}{2} \delta^{\mu\nu\rho}_{\alpha\beta\sigma} \partial_\mu h\indices{_\nu^\gamma} \partial^\alpha h\indices{^\beta_\rho}
\end{equation}
and is invariant under the gauge transformations
\begin{equation}
\delta h_{\mu\nu} = \partial_{(\mu} \xi_{\nu)}.
\end{equation}
The conjugate momenta are given by
\begin{align} \label{pi}
\pi_{ij} &= \frac{\partial \mathcal{L}_{PF}}{\partial \dot{h}^{ij}} = \frac{1}{2} \left( \dot{h}^{ij} - \delta_{ij} \dot{h} \right) - \partial_{(i} n_{j)} + \delta_{ij} \partial_k n^k .
\end{align}
The momenta conjugate to $n_i= h_{0i}$ and $N=h_{00}$ vanish identically. The trace $\pi = \pi\indices{^i_i}$ is $\pi = \frac{d-1}{2} \left( - \dot{h} + 2 \partial_k n^k \right)$, where $d$ is the space dimension.
From this, the relation \eqref{pi} can be inverted to get
\begin{equation}
\dot{h}_{ij} = 2 \left( \pi_{ij} + \partial_{(i} n_{j)} - \delta_{ij} \frac{\pi}{d-1} \right).
\end{equation}
The canonical Hamiltonian density is then
\begin{equation}
\mathcal{H}^\text{can} = \pi_{ij} \dot{h}^{ij} - \mathcal{L}_\text{PF} = \mathcal{H} + 2 n_i \mathcal{C}^i + \frac{1}{2} N \mathcal{C},
\end{equation}
where the Hamiltonian density is
\begin{align}
\mathcal{H} = \mathcal{H}_\pi + \mathcal{H}_h, \quad \mathcal{H}_\pi = \pi_{ij} \pi^{ij} - \frac{\pi^2}{d-1}, \quad \mathcal{H}_h = \frac{3}{2} \delta^{ijk}_{abc} \, \partial_i h\indices{_j^c} \partial^a h\indices{^b_k}
\end{align}
and the constraints are
\begin{align}
\mathcal{C} &= \partial_i \partial_j h^{ij} - \lap h, \\
\mathcal{C}^i &= - \partial_j \pi^{ij} .
\end{align}
Finally, the Hamiltonian action $S_H = \int\! dt \, d^d\! x\, ( \pi_{ij} \dot{h}^{ij} - \mathcal{H}^\text{can} )$ is
\begin{equation} \label{eq:hamgraviton}
S_H[h_{ij},\pi^{ij},n_i,N] = \int \! dt \, d^d\! x \left( \pi_{ij} \dot{h}^{ij} - \mathcal{H} - 2 n_i \mathcal{C}^i - \frac{1}{2} N \mathcal{C} \right).
\end{equation}
The dynamical variables are the space components $h_{ij}$ and their conjugate momenta $\pi^{ij}$. The other components of $h_{\mu\nu}$, namely $n_i = h_{0i}$ and $N = h_{00}$, only appear as Lagrange multipliers for the constraints $\mathcal{C}^i = 0$ (momentum constraint) and $\mathcal{C} = 0$ (Hamiltonian constraint).

\subsection{Hamiltonian constraint}

Up to a gauge transformation $\delta h_{ij} = 2 \partial_{(i} \xi_{j)}$, the constraint $\mathcal{C} = 0$ is solved by
\begin{equation}\label{eq:hprep}
h_{ij} = 2\,\varepsilon_{l_1 \dots l_{d-2} k (i} \partial^k \phi\indices{^{l_1 \dots l_{d-2}}_{j)}},
\end{equation}
where $\phi$ is a mixed symmetry field with the symmetries of the two-column $(d-2,1)$ Young tableau\footnote{\label{footnote:sign} Note that in $D=5$, this definition differs by a sign from \cite{Bunster:2013oaa}. The choice we make here is more convenient in arbitrary dimension.}, as is easily checked by direct substitution. The prepotential $\phi$ is determined up to a gauge and Weyl transformation of the form \eqref{eq:phigaugeapp}. Indeed, when it is plugged in equation \eqref{eq:hprep}, the second term of \eqref{eq:phigaugeapp} reproduces the gauge transformation of $h_{ij}$ with gauge parameter
\begin{equation}
\xi_i = \varepsilon_{l_1 \dots l_{d-2} k i} \partial^k A^{l_1 \dots l_{d-2}}
\end{equation}
and the other terms drop out. Remark also that the expression \eqref{eq:hprep} is traceless because of the symmetry of $\phi$ (the trace of the graviton is pure gauge).

Equation \eqref{eq:hprep} implies that the spatial components of the linearized Riemann tensor are given in terms of $\phi$ by
\begin{equation}\label{eq:Rofphi}
R\indices{^{ij}_{kl}}[h[\phi]] = \partial^{[i} \partial_{[k} h\indices{^{j]}_{l]}}[\phi] = - \frac{1}{2} \frac{1}{(d-2)!^2} \,\varepsilon^{ija_1 \dots a_{d-2}} \varepsilon_{klb_1\dots b_{d-2}} D\indices{_{a_1 \dots a_{d-2}}^{b_1 \dots b_{d-2}}}[\phi] .
\end{equation}
Using the properties of the previous section, it is easily proved that formula \eqref{eq:hprep} is unique (up to overall factors and gauge transformations). Let us spell out the proof:
\begin{enumerate}
\item The constraint $\mathcal{C} = 0$ is equivalent to the tracelessness condition
\begin{equation}
R\indices{^{ij}_{ij}} = 0
\end{equation}
on the linearized Riemann tensor of $h$. We can dualize this Riemann tensor on both groups of indices to get a tensor with $(d-2,\, d-2)$ Young symmetry (because $R$ itself has the $(2,2)$ symmetry). The constraint $\mathcal{C} = 0$ is then equivalent to the complete tracelessness of this tensor; it must therefore be equal to the Cotton tensor of some $\phi$. Dualizing back to $R$ (and adjusting factors), this is equation \eqref{eq:Rofphi}.
\item We have established that the relation \eqref{eq:Rofphi} between the Riemann tensor of $h[\phi]$ and the Cotton tensor of $\phi$ must hold. Because of gauge invariance, any expression $h_{ij} = h_{ij}[\phi]$ for $h_{ij}$ differs from \eqref{eq:hprep} only by a gauge transformation of $h_{ij}$, i.e., a term of the form $\pd_{(i} \xi_{j)}$. Moreover, invariance of the Cotton tensor implies that a gauge and Weyl transformation of $\phi$ must induce a gauge variation of $h_{ij}$.
\end{enumerate}

\subsection{Momentum constraint}

The resolution of the momentum constraint $\mathcal{C}^i = 0$ is straightforward. It gives
\begin{equation}
\pi_{ij} = \varepsilon_{ikm_1 \dots m_{d-2}} \varepsilon_{jln_1 \dots n_{d-2}} \partial^k \partial^l P^{m_1 \dots m_{d-2}  n_1 \dots n_{d-2} } \equiv G_{ij}[P],
\end{equation}
where the $(d-2\, ,d-2)$ field $P$ is determined up to the gauge and Weyl symmetries \eqref{eq:Pgaugeapp}. This follows from the construction of the Einstein tensor of $P$ carried out in the previous section.

\subsection{Prepotential action}
\label{sec:actiongraviton}

We can now plug back these solutions in the Hamiltonian action \eqref{eq:hamgraviton}. The dynamical variables $h_{ij}$ and $\pi^{ij}$ are expressed in terms of their prepotentials $\phi$ and $P$ respectively. The constraints vanish identically, so the Lagrange multipliers $n$ and $N$ disappear from the action.

The kinetic term is
\begin{align}
\pi_{ij} \,\dot{h}^{ij} &= 2 \,G_{ij}[P]\, \varepsilon^{l_1 \dots l_{d-2} ki} \partial_k \phi\indices{_{l_1 \dots l_{d-2}}^j} .
\end{align}
Integrating by parts, this can be written as
\begin{align}
\pi_{ij} \dot{h}^{ij} &= - 2 D_{i_1 \dots i_{d-2}j}[P] \,\dot{\phi}^{i_1 \dots i_{d-2}j} \\
&= 2 D_{i_1 \dots i_{d-2}j_1 \dots j_{d-2}}[\phi] \, \dot{P}^{i_1 \dots i_{d-2}j_1 \dots j_{d-2}} \label{eq:kineticPDphi} \\
&= D_{i_1 \dots i_{d-2}j_1 \dots j_{d-2}}[\phi] \, \dot{P}^{i_1 \dots i_{d-2}j_1 \dots j_{d-2}} - D_{i_1 \dots i_{d-2}j}[P] \,\dot{\phi}^{i_1 \dots i_{d-2}j}  .
\end{align}
For the Hamiltonian terms, we find
\begin{align}
\cH_\pi &= G_{ij}[P] S^{ij}[P] \\
&= P_{i_1 \dots i_{d-2}j_1 \dots j_{d-2}}\, \varepsilon^{i_1 \dots i_{d-2} kl} \partial_k D\indices{^{j_1 \dots j_{d-2}}_l}[P]
\end{align}
and
\begin{align}
\cH_h &= \frac{1}{(d-2)!} G_{i_1 \dots i_{d-2}j}[\phi] \, S^{i_1 \dots i_{d-2}j}[\phi] \\
&= \frac{1}{(d-2)!} \phi_{i_1 \dots i_{d-2}j} \, \varepsilon^{l_1 \dots l_{d-2}jk} \partial_k D\indices{^{i_1 \dots i_{d-2}}_{l_1 \dots l_{d-2}}}[\phi] ?
\end{align}
again up to total derivatives.

All in all, the action for linearized gravity becomes
\begin{align}
S[\phi, P] = \int \dtdx \,\Big( &D_{i_1 \dots i_{d-2}j_1 \dots j_{d-2}}[\phi] \, \dot{P}^{i_1 \dots i_{d-2}j_1 \dots j_{d-2}} - D_{i_1 \dots i_{d-2}j}[P] \,\dot{\phi}^{i_1 \dots i_{d-2}j} \nn \\
& - P_{i_1 \dots i_{d-2}j_1 \dots j_{d-2}}\, \varepsilon^{i_1 \dots i_{d-2} kl} \partial_k D\indices{^{j_1 \dots j_{d-2}}_l}[P] \\
& - \frac{1}{(d-2)!} \phi_{i_1 \dots i_{d-2}j} \, \varepsilon^{l_1 \dots l_{d-2}jk} \partial_k D\indices{^{i_1 \dots i_{d-2}}_{l_1 \dots l_{d-2}}}[\phi] \Big) . \nn
\end{align}
It is manifestly invariant under gauge and Weyl transformations of the prepotentials.

\subsection{Equations of motion}

The equations of motion coming from the variation of $P$ are
\begin{equation}\label{eq:Ddotphi}
\dot{D}^{i_1 \dots i_{d-2}|j_1 \dots j_{d-2}}[\phi] = - \varepsilon^{i_1 \dots i_{d-2} kl} \partial_k D\indices{^{j_1 \dots j_{d-2}}_l}[P] ,
\end{equation}
while the variation of $\phi$ yields the equation
\begin{equation}\label{eq:DdotP}
\dot{D}_{i_1 \dots i_{d-2}|j}[P] = \varepsilon_{l_1 \dots l_{d-2}jk} \partial^k D\indices{_{i_1 \dots i_{d-2}}^{l_1 \dots l_{d-2}}}[\phi] .
\end{equation}
They are equivalent to the linearized Einstein equations written in their twisted self-duality form. The proof of this fact for $D=4$ and $5$ appears in \cite{Bunster:2012km,Bunster:2013oaa}. It is done there already with the ``Cotton tensor technology"; therefore, the generalization of their proof to arbitrary $D$ is direct and we will not repeat it here.

\section{Gravitino}
\label{sec:gravitino}

In this section, we rewrite the action of the free Rarita-Schwinger field (gravitino) in terms of an antisymmetric rank  $d-2$ tensor-spinor, which we also call the ``prepotential" of the gravitino. This generalizes the result of \cite{Bunster:2012jp} to dimensions different that four, with the improvement of writing the action in terms of the appropriate Cotton tensor of section \ref{app:conffermion}. This has the advantage of making the Weyl invariance of this action more transparent.

\subsection{Hamiltonian action}

The action is
\begin{equation}\label{eq:gravitinoactioncov}
S = - \int \!d^D\!x\, \bar{\psi}_\mu \gamma^{\mu\nu\rho} \partial_\nu \psi_\rho,
\end{equation}
where the Dirac adjoint is $\bar{\psi}_\mu = i \psi_\mu^\dagger \gamma^0$. It is invariant under the gauge transformation
\begin{equation}
\delta \psi_\mu = \partial_\mu \nu
\end{equation}
where $\nu$ is an arbitrary spinor.

This action is already in first-order (Hamiltonian) form: splitting space and time indices, it is equal to
\begin{equation} \label{eq:hamgravitino}
S_H = \int \!dt\,d^d\!x\, \left( \eta^i \dot{\psi}_i - \mathcal{H} - \psi_0^\dagger \mathcal{D} - \mathcal{D}^\dagger \psi_0 \right)
\end{equation}
where the conjugate momentum, the Hamiltonian density and the constraint are
\begin{align}
\eta^i &= - i \psi^\dagger_j \gamma^{ji} \\
\mathcal{H} &= \bar{\psi}_i \gamma^{ijk} \partial_j \psi_k \\
\mathcal{D} &= -i \gamma^{ij} \partial_i \psi_j
\end{align}
respectively.
The momentum conjugate to $\psi_0$ identically vanishes, and $\psi_0$ only appears as a Lagrange multiplier for the constraint $\mathcal{D} = 0$.

\subsection{Solving the constraint}

The constraint $\mathcal{D} = 0$ is equivalent to
\begin{equation}
\partial_i \left( \gamma^{ij} \psi_j \right) = 0 .
\end{equation}
Application of the standard Poincaré lemma (with a spectator spinor index) then gives
\begin{equation}
\gamma^{ij} \psi_j = \varepsilon^{ik l_1 \dots l_{d-2}} \partial_k \chi_{l_1 \dots l_{d-2}}
\end{equation}
for some fermionic field $\chi_{l_1 \dots l_{d-2}}$ with $d-2$ antisymmetric indices.
Using the gamma matrix identity \eqref{eq:gammadelta}, we then get $\psi_i$ as
\begin{equation}\label{eq:psichi}
\psi_i = \frac{1}{d-1} \left( \gamma_{ij} - (d-2) \delta_{ij} \right) \varepsilon^{jk l_1 \dots l_{d-2}} \partial_k \chi_{l_1 \dots l_{d-2}} \equiv S_i[\chi] .
\end{equation}
where we recognize the Schouten tensor of $\chi$ as defined in section \ref{app:conffermion}. This expression reproduces the result of \cite{Bunster:2012jp} for $d=3$ (we also remark that this result was already obtained for $d=4$, albeit in a different context and in Lorentzian rather than Euclidean dimension, in the early work \cite{Deser:1980hy}). Again, $\chi$ is called the ``prepotential" of the gravitino field. It is determined up to the local gauge and Weyl transformations \eqref{eq:chigaugeapp}. Indeed, as discussed in section \ref{app:conffermion}, the transformation properties of the Schouten are exactly such that a gauge and Weyl transformation of $\chi$ produces a gauge transformation of $\psi$.

\subsection{Prepotential action}

We can now plug back the prepotential expression $\psi_i = S_i[\chi]$ into the action \eqref{eq:hamgravitino}. The kinetic term is
\begin{equation}
-i S^\dagger_i[\chi] \gamma^{ij} \dot{S}_j[\chi] = - i \chi^\dagger_{i_1 \dots i_{d-2}} \dot{D}^{i_1 \dots i_{d-2}}[\chi],
\end{equation}
where $D_{i_1 \dots i_{d-2}}[\chi]$ is the Cotton tensor of $\chi$ defined in \eqref{eq:defcottonchi} and equality holds up to a total derivative. The Hamiltonian is
\begin{align}
\bar{S}_i [\chi] \gamma^{ijk} \partial_j S_k[\chi] &= -i \, \frac{(-i)^{m+1}}{(d-3)!} G_i^\dagger[\chi]\, \gamma_{l_1 \dots l_{d-3}} \hat{\gamma} \,D^{il_1\dots l_{d-3}}[\chi] \\
&= -i \, \frac{(-i)^{m+1}}{(d-3)!} \, \chi_{i_1 \dots i_{d-2}}^\dagger \varepsilon^{i_1 \dots i_{d-2} jk} \gamma^{l_1 \dots l_{d-3}} \hat{\gamma}\, \partial_j D_{k l_1 \dots l_{d-3}}[\chi] ,
\end{align}
where $m = \lfloor D/2 \rfloor = \lfloor (d+1)/2 \rfloor$ and we used identity \eqref{eq:gijk} and integration by parts. The constraint $\mathcal{D}$ and its Lagrange multiplier $\psi_0$ disappear, since now $\mathcal{D} = 0$ identically. Putting things together, the action is
\begin{equation}\label{eq:actionchi}
S[\chi] = -i \int \!dt\,d^d\!x\, \chi_{i_1 \dots i_{d-2}}^\dagger \left( \dot{D}^{i_1 \dots i_{d-2}}[\chi] - \frac{(-i)^{m+1}}{(d-3)!} \varepsilon^{i_1 \dots i_{d-2} jk} \gamma^{l_1 \dots l_{d-3}} \hat{\gamma}\, \partial_j D_{k l_1 \dots l_{d-3}}[\chi] \right) .
\end{equation}
It can also be written in a slightly more aesthetic way as
\begin{equation}\label{eq:actiongravitinotilde}
S[\chi] = -i \int \!dt\,d^d\!x\, \chi_{i_1 \dots i_{d-2}}^\dagger \left( \dot{D}^{i_1 \dots i_{d-2}}[\chi] - \varepsilon^{i_1 \dots i_{d-2} jk} \partial_j \tilde{D}_{k}[\chi] \right) ,
\end{equation}
where we define $\tilde{D}_i[\chi]$ as the contraction
\begin{equation}
\tilde{D}_i[\chi] = \frac{(-i)^{m+1}}{(d-3)!} \,\gamma^{l_1 \dots l_{d-3}} \hat{\gamma}\, D_{i l_1 \dots l_{d-3}}[\chi],
\end{equation}
which fulfills $\gamma^i \tilde{D}_i = 0$ and $\partial^i \tilde{D}_i = 0$.


\subsection{Equations of motion}

The equations of motion following from the action \eqref{eq:actiongravitinotilde} are
\begin{equation}\label{eq:eomchi}
\dot{D}^{i_1 \dots i_{d-2}}[\chi] = \varepsilon^{i_1 \dots i_{d-2} jk} \partial_j \tilde{D}_{k}[\chi] .
\end{equation}
It is interesting to show explicitly the equivalence between this equation and the Rarita-Schwinger equation, even though it was already proven in the previous sections at the level of the action.

The Rarita-Schwinger equation is
\begin{equation} \label{eomgravitino}
\gamma^{\mu\nu\rho} F_{\nu\rho} = 0 ,
\end{equation}
where $F_{\mu\nu} = 2 \partial_{[\mu} \psi_{\nu]}$ is the field strength of the gravitino.
Taking $\mu = 0$ and $\mu = i$, it is is equivalent to the two equations
\begin{equation} \label{RSsplit}
\gamma^{ij} F_{ij} = 0, \quad \gamma^{ijk} F_{jk} - 2 \gamma^0 \gamma^{ij} F_{0j} = 0 .
\end{equation}
The first of those is the constraint $\mathcal{D} = 0$, and the second is the dynamical equation.

On the other hand, it follows from definition \eqref{eq:psichi} that the Cotton tensor is related to the field strength $F_{ij}$ as
\begin{equation}\label{eq:DofF}
D_{i_1 \dots i_{d-2}}[\chi] = \frac{1}{2}\varepsilon_{i_1 \dots i_{d-2} j k} F^{jk}, \quad\tilde{D}_i = \frac{1}{2} \gamma^0 \gamma_{ijk} F^{jk} .
\end{equation}
Using these relations in \eqref{eq:eomchi}, one finds that \eqref{eq:eomchi} is equivalent to the equations
\begin{equation}\label{eq:eomchiprime}
\dot{F}_{ij} = \gamma^0 \partial_{[i} \gamma_{j]kl} F^{kl}, \quad \gamma^{ij} F_{ij} = 0.
\end{equation}
One must keep in mind that $F_{ij}$ is expressed in terms of the prepotential $\chi$ in \eqref{eq:DofF}: this is equivalent to the constraint $\gamma^{ij} F_{ij} = 0$, which therefore must supplement the first of \eqref{eq:eomchiprime} if we write it in terms of $\psi_\mu$.

Instead of proving \eqref{eq:eomchi} $\Leftrightarrow$ \eqref{eomgravitino} directly, we will prove that \eqref{RSsplit} is equivalent to \eqref{eq:eomchiprime}. Notice that \eqref{RSsplit} contains the time components $\psi_0$ of the field $\psi_\mu$, while \eqref{eq:eomchiprime} does not but has one more derivative. This is not a problem but the usual feature of the prepotential formalism: the extra components $\psi_0$ come from \eqref{eq:eomchiprime} using the Poincaré lemma.

\begin{itemize}
\item \eqref{RSsplit} $\Rightarrow$ \eqref{eq:eomchiprime}:

Contracting the second equation with $\gamma_i$ and using the constraint, the Rarita-Schwinger equation also implies
\begin{equation}\label{eq:electricconstraint}
\gamma^i F_{0i} = 0 .
\end{equation}
(Another, equivalent way to show this is by contracting the covariant equation \eqref{eomgravitino} with $\gamma_\mu$: this gives $0 = \gamma^{\nu\rho} F_{\nu\rho} = \gamma^{ij} F_{ij} + 2 \gamma^0 \gamma^i F_{0i}$, which implies \eqref{eq:electricconstraint} using the constraint.)
This then gives the identity
\begin{equation}
\gamma^{ij} F_{0j} = ( \gamma^i \gamma^j - \delta^{ij} ) F_{0j} = - F\indices{_0^i} .
\end{equation}
Using this in the dynamical equation and multiplying by $\gamma^0$, we get
\begin{equation}
\gamma^0\gamma^{ijk} F_{jk} = 2 F\indices{_0^i}.
\end{equation}
This equation still contains the time component $\psi_0$, while \eqref{eq:eomchi} does not; we can get rid of them by taking an extra curl,
\begin{equation}
\gamma^0 \partial^{[i} \gamma^{j]kl} F_{kl} = 2 \partial^{[i} F\indices{_0^{j]}} .
\end{equation}
The right-hand side is equal to $2 \partial^{[i} \dot{\psi}^{j]} = \dot{F}^{ij}$, which then proves \eqref{eq:eomchiprime}.

\item \eqref{eq:eomchiprime} $\Rightarrow$ \eqref{RSsplit}:

Using $\dot{F}^{ij} = 2 \partial^{[i} \dot{\psi}^{j]}$, the first equation of \eqref{eq:eomchiprime} becomes
\begin{equation}
\partial_{[i} ( 2 \dot{\psi}_{j]} ) = \partial_{[i} (\gamma^0 \gamma_{j]kl} F^{kl} ).
\end{equation}
The usual Poincaré lemma then implies that $2 \dot{\psi}_{j} = \gamma^0 \gamma_{jkl} F^{kl} + 2 \partial_j \lambda$ for some spinor $\lambda$ that we are free to call $\psi_0$. This gives
\begin{equation}\label{eq:nearlyRS}
2F_{0j} = \gamma^0 \gamma_{jkl} F^{kl} .
\end{equation}
Now, contracting this equation with $\gamma^j$ and using the constraint, we get $\gamma^j F_{0j} = 0$; this in turn implies
\begin{equation}
F_{0j} = ( \delta_{ij} - \gamma_j\gamma_i )F\indices{_0^i} = \gamma_{ij} F\indices{_0^i} .
\end{equation}
We then recover \eqref{RSsplit} by putting this back in \eqref{eq:nearlyRS} and multiplying by $\gamma^0$.
\end{itemize}

\section{Supersymmetry transformations}

\label{sec:susy}

The sum of the actions \eqref{LPF} and \eqref{eq:gravitinoactioncov} is invariant under the rigid supersymmetry transformations
\begin{align}
\delta h_{\mu\nu} &= \frac{1}{2}\, \bar{\epsilon}\,\gamma_{(\mu} \psi_{\nu )} + \text{h.c.} = \frac{1}{2}\, \bar{\epsilon}\,\gamma_{(\mu} \psi_{\nu )} - 
\frac{1}{2}\, \bar{\psi}_{(\mu }\gamma_{\nu )}\epsilon,
\\
\delta \psi_{\mu} &= \frac{1}{4}\, \partial_{\rho} h_{ \mu\nu}\, \gamma^{\nu\rho}\epsilon ,
\end{align}
where $\epsilon$ is a constant spinor parameter.
In this section, we prove that the corresponding variations of the prepotentials are
\begin{align}
\delta \phi\indices{_{l_1 \dots l_{d-2} j}} &=
\frac{1}{4}\,\mathbb{P}_{(d-2,1)} \left( \bar{\epsilon} \gamma_j \chi_{l_1 \dots l_{d-2}} \right) + \hc \label{eq:deltaSUSYphi} \\
\delta P_{i_1 \dots i_{d-2} j_1 \dots j_{d-2}} &= - \frac{i^{m+1}}{4(d-2)!} \, \mathbb{P}_{(d-2, d-2)} \left( \bar{\epsilon} \, \hat{\gamma} \,\gamma_{j_{d-2} \dots j_1} \chi_{i_1 \dots i_{d-2}} \right) + \hc \label{eq:deltaSUSYP}
\end{align}
for the bosonic prepotentials and $\delta \chi = \delta_\phi \chi + \delta_P \chi$, with
\begin{align}
\delta_\phi \chi_{i_1 \dots i_{d-2}} = - \frac{(-i)^{m+1}}{2(d-2)!} \bigg[ &\varepsilon_{jkl_1 \dots l_{d-2}} \pd^j \phi\indices{_{i_1 \dots i_{d-2}}^k} \gamma^{l_1 \dots l_{d-2}} \label{eq:deltaSUSYphichi} \\
&+ \frac{(d-2)(d-3)}{2}\varepsilon_{[i_1|jk_1 \dots k_{d-2}|} \pd^j \phi\indices{^{k_1 \dots k_{d-2}}_{i_2}} \gamma_{i_3 \dots i_{d-2}]} \bigg] \hat{\gamma} \gamma^0 \epsilon \nn
\end{align}
and
\begin{equation}
\delta_P \chi_{i_1 \dots i_{d-2}} = - \frac{1}{2} \partial^p P\indices{_{i_1 \dots i_{d-2}}^{q_1 \dots q_{d-2}}}\varepsilon_{jpq_1 \dots q_{d-2}} \gamma^j \gamma^0 \epsilon \label{eq:deltaSUSYPchi}
\end{equation}
for the fermionic one. This is the generalization of the transformations found in \cite{Bunster:2012jp} for $d=3$ to arbitrary dimension. (Note that the second line of \eqref{eq:deltaSUSYphichi} is absent in $d=3$.)

\subsection{Variation of the first bosonic prepotential $\phi$}

The spatial components of the covariant expression give
\begin{equation}\label{eq:deltahij}
\delta h_{ij} = \frac{1}{2} \bar{\epsilon}\gamma_{(i} \psi_{j )} + \hc.
\end{equation}
By expressing $\psi_i$ in terms of its prepotential, one gets
\begin{align}
\delta h_{ij} &= \frac{1}{2} \bar{\epsilon}\gamma_{(i} \psi_{j)} + \hc \\
&= \frac{1}{2(d-1)} \bar{\epsilon}\gamma_{(i} \left( \gamma_{j)k} - (d-2) \delta_{j)k} \right) \varepsilon^{km l_1 \dots l_{d-2}} \partial_m \chi_{l_1 \dots l_{d-2}} + \hc \\
&= \frac{1}{2(d-1)} \bar{\epsilon} \left( \delta_{ij} \gamma_{k} - (d-1) \delta_{k(j}\gamma_{i)} \right) \varepsilon^{km l_1 \dots l_{d-2}} \partial_m \chi_{l_1 \dots l_{d-2}} + \hc \\
&= \frac{1}{2(d-1)} \bar{\epsilon} \partial_m \left( \delta_{ij} \gamma_{k}\varepsilon^{km l_1 \dots l_{d-2}} \chi_{l_1 \dots l_{d-2}} 
- (d-1) \gamma_{(i}\varepsilon_{j)}^{\phantom{j)}m l_1 \dots l_{d-2}} \chi_{l_1 \dots l_{d-2}}  \right) + \hc .
\end{align}
By noting that a full antisymmetrization over all the indices of the $\epsilon$ and the $j$ of the $\delta$ vanishes identically, we can rewrite the first term as
\begin{align}
\bar{\epsilon} \partial^m  \delta_{ij} \gamma^{k}\varepsilon_{km l_1 \dots l_{d-2}} \chi^{l_1 \dots l_{d-2}} 
&= \bar{\epsilon} \partial^m  \delta_{ik} \gamma^{k}\varepsilon_{jm l_1 \dots l_{d-2}} \chi^{l_1 \dots l_{d-2}} \nn \\
&\quad + \bar{\epsilon} \partial^m  \delta_{im} \gamma^{k}\varepsilon_{kj l_1 \dots l_{d-2}} \chi^{l_1 \dots l_{d-2}} \nn \\
&\quad + \left(d-2\right)\bar{\epsilon} \partial^m  \delta_{il_1} \gamma^{k}\varepsilon_{km j l_2 \dots l_{d-2}} \chi^{l_1 \dots l_{d-2}} \\ &= \bar{\epsilon} \partial^m  \gamma_{i}\varepsilon_{jm l_1 \dots l_{d-2}} \chi^{l_1 \dots l_{d-2}} \nn \\
&\quad + \bar{\epsilon} \partial_i \gamma^{k}\varepsilon_{kj l_1 \dots l_{d-2}} \chi^{l_1 \dots l_{d-2}} \nn \\
&\quad + \left(d-2\right)\bar{\epsilon} \partial^m \gamma^{k}\varepsilon_{km j l_2 \dots l_{d-2}} \chi_i^{\ l_2 \dots l_{d-2}} .
\end{align}
Since this expression is already symmetric in $ij$, we can explicitly symmetrize it again. The second term then appears as a gauge variation of $h_{ij}$ and can be discarded. We then obtain up to a gauge transformation
\begin{equation} \label{eq:varhij}
\delta h_{ij} = \frac{d-2}{2(d-1)} \bar{\epsilon} \partial_m \left( 
 \gamma_{k}\varepsilon_{(i}^{\phantom{(i} km l_2 \dots l_{d-2}} \chi_{j) l_2 \dots l_{d-2}}
-  \gamma_{(i}\varepsilon_{j)}^{\phantom{j)}m l_1 \dots l_{d-2}} \chi_{l_1 \dots l_{d-2}} \right) + \hc.
\end{equation}
On the other hand, carrying out the projection in \eqref{eq:deltaSUSYphi},
\begin{align}
\delta \phi\indices{_{l_1 \dots l_{d-2}\vert j}}
&= \frac{1}{4}\,\mathbb{P}_{(d-2,1)} \left( \bar{\epsilon} \gamma_j \chi_{l_1 \dots l_{d-2}} \right) + \hc \nn \\
&= \frac{1}{4} \left[\bar{\epsilon} \gamma_j \chi_{l_1 \dots l_{d-2}} - \bar{\epsilon} \gamma_{[j} \chi_{l_1 \dots l_{d-2}]}\right] + \hc \nn \\
&= \frac{d-2}{4(d-1)}\left[\bar{\epsilon} \gamma_j \chi_{l_1 \dots l_{d-2}} + \bar{\epsilon}\gamma_{[l_1} \chi_{\vert j \vert l_2 \dots l_{d-2} ] }\right] + \hc,
\end{align}
our ansatz gives
\begin{align}
\delta h_{ij} &= 2\,\varepsilon^{l_1 \dots l_{d-2}}_{\phantom{l_1 \dots l_{d-2}} k(i\vert} \, \partial^k \delta \phi_{l_1 \dots l_{d-2}\vert j)} \nn \\
&= \frac{d-2}{2(d-1)} \bar{\epsilon} \partial_m \left[\varepsilon^{l_1 \dots l_{d-2}m}_{\phantom{l_1 \dots l_{d-2}m} (i\vert} \,  \gamma_{\vert j)} \chi_{l_1 \dots l_{d-2}}
+ \varepsilon^{k l_2 \dots l_{d-2}m}_{\phantom{kl_1 \dots l_{d-2}m} (i\vert} \, \gamma_{k} \chi_{\vert j) l_2 \dots l_{d-2} } \right] ,
\end{align}
which is the same expression as \eqref{eq:varhij}.

\subsection{Variations of the fermionic prepotential}

Splitting time and space, one has $\delta \psi_i = \delta_h \psi_i + \delta_\pi \psi_i$ for the spatial components of the gravitino, with
\begin{align}
\delta_h \psi_i &= \frac{1}{4} \partial_k h_{ij} \gamma^{jk} \epsilon, \\
\delta_\pi \psi_i &= \frac{1}{2} \left( \pi_{ij} - \delta_{ij} \frac{\pi}{d-1} \right) \gamma^j \gamma^0 \epsilon .
\end{align}
Expressing this in terms of prepotentials, it follows that the variation $\delta \chi = \delta_\phi \chi + \delta_P \chi$ of the fermionic prepotential must be such that
\begin{equation}\label{eq:schoutensusy}
S_i[\delta_\phi \chi] = \frac{1}{4} \partial_k h_{ij}[\phi] \gamma^{jk} \epsilon, \quad S_i[\delta_P \chi] = \frac{1}{2} S_{ij}[P] \gamma^j \gamma^0 \epsilon .
\end{equation}

\paragraph{Variation $\delta_\phi \chi$ containing the first prepotential.}

First, it follows from \eqref{eq:schoutensusy} that the variation of the Cotton tensor of $\chi$ is
\begin{align}
D_{i_1 \dots i_{d-2}}[\delta_\phi \chi] &= - \frac{1}{4} \varepsilon_{i_1 \dots i_{d-2}kl} R\indices{^{kl}_{pq}}[h[\phi]] \gamma^{pq} \epsilon \\
&= \frac{1}{4(d-2)!} D\indices{_{i_1 \dots i_{d-2}}^{j_1 \dots j_{d-2}}}[\phi] \varepsilon_{j_1 \dots j_{d-2} pq} \gamma^{pq} \epsilon, \label{eq:deltachiD}
\end{align}
where we used the relation \eqref{eq:Rofphi} between the Riemann of $h[\phi]$ and the Cotton of $\phi$. Writing both sides as the curl of the respective Schouten tensors, this gives
\begin{equation}
S_i[\delta_\phi \chi] = \frac{1}{4(d-2)!} S\indices{^{j_1 \dots j_{d-2}}_i} [\phi] \varepsilon_{j_1 \dots j_{d-2} kl} \gamma^{kl} \epsilon
\end{equation}
up to a total derivative that can cancelled by adding the appropriate Weyl transformation to $\delta_\phi \chi$. (This is more convenient than \eqref{eq:schoutensusy} because it involves directly the Schouten tensors on both sides.) We now write this in terms of the Einstein tensors: this gives
\begin{equation}
G_i[\delta_\phi \chi] = - \frac{(-i)^{m+1}}{2(d-2)!}\left( G_{j_1 \dots j_{d-2} i}[\phi] \gamma^{j_1 \dots j_{d-2}}  + \frac{(d-2)(d-3)}{2} G\indices{_{k i j_3 \dots j_{d-2}}^k} [\phi] \gamma^{j_3 \dots j_{d-2}} \right) \hat{\gamma} \gamma^0 \epsilon .
\end{equation}
The computation uses relations \eqref{eq:GofSchi}, \eqref{eq:SofGphi} and the gamma matrix identity \eqref{eq:g2gd-2}. From there, the variation $\delta_\phi \chi$ is found by writing the right-hand side as a curl: this yields expression \eqref{eq:deltaSUSYphichi}.

\paragraph{Variation $\delta_P \chi$ containing the second prepotential.}

Using the identity $\gamma_{ij} S^j[\chi] = G_i[\chi]$, we must have for the Einstein tensor
\begin{equation}
G_i[\delta_P \chi] = \frac{1}{2} S\indices{^j_k}[P] \gamma_{ij} \gamma^k \gamma^0 \epsilon .
\end{equation}
The variations $\delta_P \chi$ can the be identified up to a gauge transformation by writing the right-hand sides as a curl. Using the identity $\gamma_{ij} \gamma^k = \gamma\indices{_{ij}^k} + \gamma_i \delta^k_j - \gamma_j \delta^k_i$ and the fact that $S_{ij}[P]$ is symmetric, we get
\begin{align}
\frac{1}{2} S\indices{^j_k}[P] \gamma_{ij} \gamma^k \gamma^0 \epsilon &= - \frac{1}{2} (S_{ij}[P] - \delta_{ij} S\indices{^k_k}[P] ) \gamma^j \gamma^0 \epsilon = - \frac{1}{2} G_{ij}[P] \gamma^j \gamma^0 \epsilon \\
&= \varepsilon_{ikl_1\dots k_{d-2}} \partial^k \left( - \frac{1}{2}\varepsilon_{jpq_1 \dots q_{d-2}} \partial^p P^{l_1 \dots l_{d-2} q_1 \dots q_{d-2}} \gamma^j \gamma^0 \epsilon \right)
\end{align}
from which we deduce the expression \eqref{eq:deltaSUSYPchi}.

\subsection{Variation of the second bosonic prepotential $P$}

We now determine the variation of $P$ from the invariance of the kinetic term of the prepotential action. Equation \eqref{eq:deltachiD} implies for the hermitian conjugate
\begin{equation}
(D_{i_1 \dots i_{d-2}}[\delta_\phi \chi])^\dagger = - \frac{i^m}{2(d-2)!} \bar{\epsilon} \, \hat{\gamma} \gamma_{j_{d-2} \dots j_1} D\indices{_{i_1 \dots i_{d-2}}^{j_1 \dots j_{d-2}}}[\phi] ,
\end{equation}
where we used the gamma matrix identity \eqref{eq:gij}. The variation of the kinetic term of $\chi$ is then readily computed. It cancels with the variation of the bosonic kinetic term (more conveniently written in the form ``$P \, \dot{D}[\phi]$'', see \eqref{eq:kineticPDphi}) provided the variation of $P$ is given by \eqref{eq:deltaSUSYP}.

Checking directly that \eqref{eq:deltaSUSYP} reproduces the supersymmetry variation of the momentum $\pi_{ij}$ is cumbersome because of the many terms contained in the projection. However, it is not necessary since, if the variations of the other fields are known, the variation of $\pi_{ij}$ is uniquely determined by the invariance of the kinetic term in the Hamiltonian action.

\section{Conclusions}

We have obtained a formulation of linearized gravity in terms of prepotentials valid in arbitrary dimension, based on a systematic use of conformal geometry, whose appropriate tools we have developed for the relevant fields. The action principle and the equations of motion of the graviton and gravitino and the supersymmetry variations rotating them into each other have been rewritten in terms of our new variables in a geometrically transparent way. The dimensional reduction has also been reproduced using the prepotentials and the consistency with the similar treatment of the maximally supersymmetric $\mathcal{N}=(4,0)$ six-dimensional theory has been confirmed.

A natural extension of these results would be to consider the interacting case. However, there is no clear path to the resolution of the non linear constraints and the persistence of duality invariance is excluded for fully interacting gravitation \cite{Deser:2005sz}. Moreover, no-go theorems exists for local self-interactions of the dual graviton (Curtright field), see \cite{Bekaert:2002uh,Bekaert:2004dz}. Notwithstanding these difficulties, see \cite{Barnich:2008ts} for an introduction of sources and \cite{Julia:2005,Julia:2005ze,Leigh:2007wf,Hortner:2016omi} for the inclusion of a cosmological constant. Let us also mention that duality was clarified for the massless, massive and ``partially-massless" gravitons in the recent preprint \cite{Boulanger:2018shp}, including arbitrary dimension and cosmological constant. The actions they use include both the graviton and dual graviton fields in a manifestly Lorentz-invariant manner, but they are not on the same footing (see also \cite{Boulanger:2003vs,Boulanger:2008nd}): it would be interesting to understand how they relate to the actions of the present paper\footnote{These methods were also used in \cite{Boulanger:2012df,Boulanger:2012mq,Boulanger:2015mka,Bergshoeff:2016ncb} to investigate the ``double dual graviton" and other exotic dualizations of $p$-forms and mixed symmetry gauge fields. Again, how this is realized in the first-order, unconstrained Hamiltonian formalism remains to be explored. We are grateful to Nicolas Boulanger for pointing out these references.}.

Duality is also closely related to $E_{10}$, which was argued \cite{Damour:2002cu} to be the infinite-dimensional symmetry of the tensionless limit of string theory. This limit contains an infinite tower of massless higher spin modes. The manifestly duality invariant rewriting of higher spin gauge fields is therefore of closely related interest and has already been the object of extensive investigation \cite{Bekaert:2003az,Boulanger:2003vs,Deser:2004xt,Henneaux:2015cda,Henneaux:2016zlu}. In the bosonic case, the free theory has been expressed in terms of prepotentials which enjoy the usual conformal invariance, the equations of motion being again equivalent to twisted self-duality conditions. With a view toward obtaining a manifestly duality invariant form of higher spin supermultiplets, an upcoming paper \cite{HLL:prep} will extend our findings to fermionic higher spins (construction of conformal curvatures, resolution of Hamiltonian constraints and rewriting of the action and equations of motion).

Four-dimensional gravitational duality arises geometrically from toroidal reduction of the six-dimensional models of \cite{Hull:2000zn,Hull:2000rr}. The construction of a self-contained action principle is a first step towards a better understanding of these exotic theories and relies on the introduction of prepotentials. This was done for the $(2,2)$ chiral tensor and its $\mathcal{N} = (4,0)$ supersymmetric extension in \cite{Henneaux:2016opm,Henneaux:2017xsb}; the upcoming paper \cite{HLMS:prep} will extend these results to the chiral $(2,1)$ tensor appearing in the $\mathcal{N} = (3,1)$ maximally supersymmetric theory.

Finally, the prepotential formalism for the gravitino field developed in this paper should provide an avenue for exploring the dual formulations of supergravity alluded to in \cite{Curtright:1980yk}. It is planned to return to this question in the future.

\section*{Acknowledgments}

We would like to thank Marc Henneaux for useful discussions during the early stages of this project and for comments on the final draft. V.L. is grateful to Nicolas Boulanger for discussions and comments, and also to Javier Matulich and Stefan Prohazka for useful remarks on this work which arose during the preparation of \cite{HLMS:prep}. A.L. is deeply thankful for the fruitful work environment offered by the Albert Einstein Institute of Gravitational Physics of Potsdam. This work was partially supported by the ERC Advanced Grant ``High-Spin-Grav''. V.L. is a Research Fellow at the Belgian F.R.S.-FNRS.

\appendix

\section{Properties of the Cotton tensors}
\label{app:cotton}

In this appendix, we prove the two important properties of the Cotton tensors presented in section \ref{sec:conformal}.

\subsection{Gauge and Weyl invariance}

The first property is that the Cotton tensor completely captures gauge and Weyl invariance.

\paragraph{Bosonic $(d-2,1)$ field.}

We want to prove the implication
\begin{equation}
D_{i_1 \dots i_{d-2} j_1 \dots j_{d-2}}[\phi] = 0 \;\Rightarrow\; \phi\indices{^{i_1 \dots i_{d-2}}_{j} } = \text{(gauge)} + \delta^{[i_1}_j B^{i_2 \dots i_{d-2}]} \;\text{ for some } B.
\end{equation}
(The opposite implication is true by construction.) The proof goes in two steps:
\begin{enumerate}
\item First, $D = 0$ implies that the curl of the Schouten tensor on both groups of indices vanishes,
\begin{equation}
\partial^{[i_1} S\indices{^{i_2 \dots i_{d-1}]}_j} = 0, \qquad \partial_{[j} S\indices{^{i_1 \dots i_{d-2}}_{k]}} = 0 .
\end{equation}
Indeed, the second of those equations follows from the definition of the Cotton tensor, and the first from the relation \eqref{eq:dprimephi} between this curl and the trace of the Cotton. The generalized Poincaré lemma of \cite{Bekaert:2002dt} then implies that $S$ is itself a double curl, i.e.
\begin{equation}
S\indices{^{i_1 \dots i_{d-2}}_{j}} [\phi] = - (d-2)!\,\partial_j\partial^{[i_1} B^{i_2 \dots i_{d-2}]}
\end{equation}
for some antisymmetric $B$. This is precisely the variation of the Schouten induced by a Weyl transformation of $\phi$.
\item Then, using the inversion relations between the Einstein and Cotton tensors, this implies that
\begin{equation}
G\indices{^{i_1 \dots i_{d-2}}_{j}} [\phi - \delta B] = \partial^k \partial_m \left( \phi\indices{^{l_1 \dots l_{d-2}}_{n} } - \delta^{[i_1}_j B^{i_2 \dots i_{d-2}]} \right) \varepsilon^{mni_1 \dots i_{d-2}} \varepsilon_{jkl_1 \dots l_{d-2}} = 0
\end{equation}
where $\delta B$ stands for the tensor $\delta^{[i_1}_j B^{i_2 \dots i_{d-2}]}$. The Poincaré lemma of \cite{Bekaert:2002dt} (property \eqref{eq:gaugeEinsteinphi} of the Einstein tensor) now implies that $\phi - \delta B$ is pure gauge, which is what we wanted to prove.
\end{enumerate}

\paragraph{Bosonic $(d-2,d-2)$ field.} The implication to be proven is now
\begin{equation}
D\indices{^{i_1 \dots i_{d-2}}_{j}}[P] = 0 \;\Rightarrow\; P\indices{^{i_1 \dots i_{d-2}}_{j_1 \dots j_{d-2}} } = \text{(gauge)} + \delta^{i_1 \dots i_{d-2}}_{j_1 \dots j_{d-2}} \,\xi \;\text{ for some } \xi.
\end{equation}
The proof goes as before:
\begin{enumerate}
\item First, $D=0$ implies that the curl of $S_{ij}[P]$ vanishes,
\begin{equation}
\partial_{[i} S_{j]k} = 0.
\end{equation}
The Poincaré lemma of \cite{DuboisViolette:1999rd,DuboisViolette:2001jk} applied to symmetric tensors implies then that
\begin{equation}
S_{ij}[P] = -(d-2)! \pd_i \pd_j \xi
\end{equation}
for some $\xi$, which is again the form induced by a Weyl transformation of $\xi$.
\item Then, this implies that the Einstein tensor of $P - \delta^{d-2} \xi$ vanishes, where $\delta^{d-2} \xi$ stands for $\delta^{i_1 \dots i_{d-2}}_{j_1 \dots j_{d-2}} \,\xi$. The combination $P - \delta^{d-2} \xi$ is therefore pure gauge, which proves the proposition.
\end{enumerate}

\paragraph{Fermionic $(d-2)$ field.} We now should prove
\begin{equation}
D_{i_1 \dots i_{d-2}}[\chi] = 0 \;\Rightarrow\; \chi_{i_1 \dots i_{d-2}} =  \text{(gauge)} + \gamma_{i_1 \dots i_{d-2}} \rho \;\text{ for some } \rho.
\end{equation}
We take the same steps:
\begin{enumerate}
\item First, $D=0$ is equivalent to $\partial_{[i} S_{j]}[\chi] = 0$, which by the usual Poincaré lemma with a spectator spinorial index implies
\begin{equation}
S_i[\chi] = \pd_i \nu
\end{equation}
for some $\nu$ that can always be written as $\nu = i^{m+1} (d-2)! \gamma_0 \hat{\gamma}\, \rho$. This is the form of the Schouten induced by a Weyl transformation of $\chi$.
\item This implies that the Einstein tensor of $\chi_{i_1 \dots i_{d-2}} - \gamma_{i_1 \dots i_{d-2}} \rho$ is zero. That combination is therefore pure gauge, which is what had to be proven.
\end{enumerate}

\subsection{Conformal Poincaré lemma}

The second property is a ``conformal version" of the Poincaré lemma: any divergenceless tensor field with the same symmetry and trace properties as the Cotton tensor can always be written as the Cotton tensor of the appropriate field.

\paragraph{Bosonic $(d-2,1)$ field.}

We need to prove that, for any $(d-2, d-2)$ tensor $T_{i_1 \dots i_{d-2} j_1 \dots j_{d-2}}$ that is divergenceless and completely traceless, there exists a $(d-2,1)$ field $\phi\indices{^{i_1 \dots i_{d-2}}_{j} }$ such that $T = D[\phi]$. Moreover, $\phi$ is determined from this condition up to gauge and Weyl transformations of the form \eqref{eq:phigaugeapp}.
\begin{enumerate}

\item First, the divergencelessness of $T$ and the usual Poincaré lemma implies that $T$ can be written as the curl of some other tensor $U$,
\begin{equation}\label{eq:TofU}
T^{i_1 \dots i_{d-2}  j_1 \dots j_{d-2}} = \varepsilon^{i_1 \dots i_{d-2}kl} \partial_k U\indices{^{j_1 \dots j_{d-2}}_{l}} .
\end{equation}
The tensor $U$ is determined up to $U\indices{^{j_1 \dots j_{d-2}}_{l}} \rightarrow U\indices{^{j_1 \dots j_{d-2}}_{l}} + \pd_l V^{j_1 \dots j_{d-2}}$, where $V$ is totally antisymmetric.

\item At this stage, $U$ is not of irreducible Young symmetry $(d-2,1)$, but could have a completely antisymmetric component. However, the condition that $T$ is completely traceless implies that this component satisfies $\pd_{[k} U_{j_1 \dots j_{d-2} l]} = 0$, i.e., is of the form $U_{[j_1 \dots j_{d-2} l]} = \pd_{[l} W_{j_1 \dots j_{d-2}]}$ for some antisymmetric $W$. Now, the ambiguity in $U$ described above can be used to precisely cancel this contribution. We can therefore assume the $U$ has the $(d-2,1)$ symmetry from now on.

\item We now use the fact that $T$ has the $(d-2,d-2)$ symmetry. This implies that
\begin{equation}
\varepsilon_{i_1 \dots i_{d-2} j_1 m} T^{i_1 \dots i_{d-2}  j_1 \dots j_{d-2}} = 0
\end{equation}
which, using the expression \eqref{eq:TofU}, is equivalent to the differential identity
\begin{equation}\label{eq:divU}
\partial_{i_1} U\indices{^{i_1 \dots i_{d-2}}_{j}} [\phi] - \partial_j U^{i_2 \dots i_{d-2}} = 0
\end{equation}
on $U$. We now define another tensor $X$ of $(d-2,1)$ symmetry by the equation
\begin{equation}
X\indices{^{i_1 \dots i_{d-2}}_{j}} = U\indices{^{i_1 \dots i_{d-2}}_{j}} - (d-2) \,\delta^{[i_1}_j U\indices{^{i_2 \dots i_{d-2}]k}_k} .
\end{equation}
The identity \eqref{eq:divU} is then equivalent to the fact that $X$ is divergenceless.

\item We are now ready to conclude. Because $X$ is a divergenceless $(d-2,1)$ tensor, it must be the Einstein tensor of some $(d-2,1)$ field $\phi$, $X = G[\phi]$. The relation between $U$ and $X$ then implies that $U$ is the Schouten tensor of $\phi$, $U = S[\phi]$, and equation \eqref{eq:TofU} gives at last $T = D[\phi]$, which was to be proven.

\item The ambiguity in $\phi$ is easily determined from the first property of the Cotton tensor. Indeed, by linearity, the condition $D[\phi'] = D[\phi]$ is equivalent to $D[\phi'-\phi] = 0$. The first property of the Cotton then implies that $\phi' - \phi$ takes the form ``Weyl $+$ gauge transformation".
\end{enumerate}

\paragraph{Bosonic $(d-2,d-2)$ field.}

For the $(d-2,d-2)$ field, we now prove that, for any divergenceless, traceless $(d-2, 1)$ tensor $T_{i_1 \dots i_{d-2} j}$, there exists a $(d-2,d-2)$ field $P_{i_1 \dots i_{d-2} j_1 \dots j_{d-2}}$ such that $T = D[P]$. This field $P$ is determined from this condition up to a gauge and Weyl transformation of the form \eqref{eq:Pgaugeapp}. The proof goes as before:

\begin{enumerate}

\item First, the divergencelessness of $T$ implies that it can be written as a curl,
\begin{equation}\label{eq:TofUP}
T^{i_1 \dots i_{d-2}  j} = \varepsilon^{i_1 \dots i_{d-2}kl} \partial_k U\indices{^{j}_{l}} .
\end{equation}
The tensor $U$ is determined up to $U\indices{^{j}_{l}} \rightarrow U\indices{^{j}_{l}} + \pd_l V^{j}$.

\item The tensor $U$ is not necessarily symmetric. However, the condition that $T$ is traceless implies that the antisymmetric component satisfies $\pd_{[k} U_{jl]} = 0$, i.e., is of the form $U_{[jl]} = \pd_{[l} W_{j]}$ for some vector $W$. By using $V$, we can precisely cancel this contribution and we can assume that $U$ is symmetric.

\item We now use the fact that $T$ satisfies $T_{[i_1 \dots i_{d-2}j]}=0$. This is equivalent to
\begin{equation}
\pd^i U_{ij} - \pd_j U\indices{_k^k} = 0.
\end{equation}
Defining the symmetric tensor $X$ by
\begin{equation}
X_{ij} = U_{ij} - \delta_{ij} U\indices{_k^k},
\end{equation}
that identity is equivalent to the fact that $X$ is divergenceless.

\item We can now conclude: because $X$ is a symmetric divergenceless tensor, it must be the Einstein tensor of some $(d-2,d-2)$ field $P$, $X = G[P]$. The relation between $U$ and $X$ then implies that $U$ is the Schouten tensor of $P$, $U = S[P]$, and equation \eqref{eq:TofUP} gives $T = D[\phi]$.

\item The ambiguity in $P$ is given by the first property of the Cotton tensor. Indeed, the condition $D[P'] = D[P]$ is equivalent to $D[P'-P] = 0$, which shows that $P'$ and $P$ differ by a gauge and Weyl transformation.
\end{enumerate}

\paragraph{Fermionic $(d-2)$ field.}

We fnish this appendix by proving the analogous property for the fermionic antisymmetric field, namely: For any divergenceless, completely gamma-traceless rank $d-2$ tensor-spinor $T_{i_1 \dots i_{d-2}}$, there exists an antisymmetric tensor-spinor $\chi_{i_1 \dots i_{d-2}}$ such that $T = D[\chi]$. Moreover, $\chi$ is determined from this condition up gauge and Weyl transformations of the form \eqref{eq:chigaugeapp}. We follow the same steps as before, but without the complications of Young symmetries:
\begin{enumerate}

\item The divergencelessness of $T$ implies that it can be written as
\begin{equation}\label{eq:TofUchi}
T^{i_1 \dots i_{d-2}} = \varepsilon^{i_1 \dots i_{d-2}kl} \partial_k U_l,
\end{equation}
where $U$ is determined up to a total derivative $\pd_l V$ for some spinor field $V$.

\item Using the gamma matrix identity \eqref{eq:gij}, the complete gamma-tracelessness
\begin{equation}
\gamma^{i_1 \dots i_{d-2}} T_{i_1 \dots i_{d-2}} = 0    
\end{equation}
of $T$ is equivalent to the differential equation $\gamma_{ij} \pd^i U^j = 0$, which is in turn equivalent to $\pd^i X_i = 0$ if we define $X_i = \gamma_{ij} S^j$.

\item Because $X$ is a divergenceless vector-spinor, it is the Einstein tensor of some antisymmetric tensor-spinor $\chi$, $X = G[\chi]$. As before, the relation between $U$ and $X$ then implies that $U$ is the Schouten tensor of $\chi$, $U = S[P]$, and the relation between $T$ and $U$ then gives $T = D[\phi]$.

\item The ambiguity in $\chi$ is given by the first property of the Cotton tensor: $D[\chi'] = D[\chi]$ is equivalent to $D[\chi'-\chi] = 0$, which shows that $\chi'$ and $\chi$ differ by a transformation of the form \eqref{eq:chigaugeapp}.
\end{enumerate}

\section{Lower-spin fields}
\label{app:lowspin}

We recall here the Hamiltonian formulation of the free fields of spin $0$, $1/2$ and $1$ in our notations. In the case of spin $1$, the constraints associated to the gauge invariance $\delta A_\mu = \partial_\mu \lambda$ can be solved, yielding an action with another $(d-2)$-form potential but no constraint \cite{Bunster:2011qp}. These are the actions that naturally appear by dimensional reduction of the graviton and gravitino actions in the prepotential formalism. For completeness, we also review the dimensional reduction of the free vector action.

\subsection{Scalar field}

The Hamiltonian action for a free scalar field is
\begin{equation} \label{eq:Hs}
S_H = \int \dtdx \left( \pi \dot{\varphi} - \mathcal{H}_s \right), \quad \mathcal{H}_s = \frac{1}{2} \pi^2 + \frac{1}{2} \partial_i \varphi \, \partial^i \varphi,
\end{equation}
where $\pi = \dot{\varphi}$ is the momentum conjugate to $\varphi$.

\subsection{Dirac field}
\label{app:dirac}

The action for a Dirac field is alrady in first-order form. It is
\begin{align}
S_{\frac{1}{2}} = - \int \dx\, \bar{\psi}\, \gamma^\mu \partial_\mu \psi = i \int \dtdx \, \psi^\dagger \left( \dot{\psi} - \gamma^0 \gamma^i \partial_i \psi \right) .
\end{align}

\subsection{Vector field}
\label{app:vector}

For a free vector field, the Hamiltonian action is
\begin{equation} \label{eq:Hv}
S_H = \int \! dt \, d^d\! x \left( \pi_i \dot{A}^i - \mathcal{H}_v - A_0 \mathcal{G} \right), \quad \mathcal{H}_v = \frac{1}{2} \pi_i \pi^i + \frac{1}{4} F_{ij} F^{ij}, \quad \mathcal{G} = - \partial_i \pi^i,
\end{equation}
where $F_{\mu\nu} = \partial_\mu A_\nu - \partial_\nu A_\mu$ is the field strength and $\pi_i = F_{0i} = \dot{A}_i - \partial_i A_0$ is the momentum conjugate to $A_i$. The time component $A_0$ only appears as a Lagrange multiplier for the constraint $\mathcal{G} = 0$ (Gauss constraint).

Following \cite{Bunster:2011qp}, the constraint $\cG = 0$ can be solved by
\begin{equation}
\pi^i = \frac{1}{(d-2)!}\varepsilon^{ijk_1 \dots k_{d-2}} \partial_j Z_{k_1 \dots k_{d-2}} = \cB^i[Z],
\end{equation}
where we defined the magnetic field of a $(d-2)$-form $Z$. We define in a similar manner
\begin{equation}
\cB^{i_1 \dots i_{d-2}}[A] = \frac{1}{2}\varepsilon^{i_1 \dots i_{d-2} jk} F_{jk} = \varepsilon^{i_1 \dots i_{d-2} jk} \partial_j A_k .
\end{equation}
The action \eqref{eq:Hv} then takes the form
\begin{equation}
S = \int \dtdx \left( K_v - \cH_v \right),
\end{equation}
where the kinetic term $K_v$ can be written in several different ways up to integration by parts,
\begin{align}
K_v &= \cB_i[Z] \dot{A}^i \nonumber \\
&= - \frac{1}{(d-2)!} \cB_{i_1 \dots i_{d-2}}[A] \dot{Z}^{i_1 \dots i_{d-2}} \nonumber\\
&= \frac{1}{2} \left( \cB_i[Z] \dot{A}^i - \frac{1}{(d-2)!} \cB_{i_1 \dots i_{d-2}}[A] \dot{Z}^{i_1 \dots i_{d-2}} \right),
\end{align}
and the Hamiltonian density is
\begin{align}
\cH_v &= \frac{1}{2} \left( \cB_i[Z] \cB^i[Z] + \frac{1}{(d-2)!} \cB_{i_1 \dots i_{d-2}}[A] \cB^{i_1 \dots i_{d-2}}[A] \right) \\
&= \frac{1}{2(d-2)!} \left( Z_{i_1 \dots i_{d-2}} \varepsilon^{i_1 \dots i_{d-2}jk} \pd_j \cB_k[Z] + A_i \varepsilon^{ijk_1 \dots k_{d-2}} \partial_j \cB_{k_1 \dots k_{d-2}}[A] \right) .
\end{align}

\paragraph{Dimensional reduction.}

We reduce from $D+1=d+2$ to $D=d+1$ dimensions and write the extra coordinate as $z$. The two higher-dimensional potentials $\hat{A}$ and $\hat{Z}$ split as
\begin{equation}
\hat{A}_I \;\rightarrow\; \hat{A}_i,\; \hat{A}_z, \quad \hat{Z}_{I_1 \dots I_{d-1}} \;\rightarrow\; \hat{Z}_{i_1 \dots i_{d-1}},\; \hat{Z}_{i_1 \dots i_{d-2} z},
\end{equation}
with gauge transformations
\begin{align}
\delta \hat{A}_i &= \partial_i \Lambda, & \delta \hat{Z}_{i_1 \dots i_{d-1}} &= (d-1) \partial_{[i_1} \tilde{\Lambda}_{i_2 \dots i_{d-1}]}, \\
\delta \hat{A}_z &= 0, & \delta \hat{Z}_{i_1 \dots i_{d-2} z} &= (d-2) \partial_{[i_1} \tilde{\Lambda}_{i_2 \dots i_{d-2}] z} .
\end{align}
The potentials $\hat{A}_i$ and $\hat{Z}_{i_1 \dots i_{d-2} z}$ are those of a vector field in $D$ dimensions, and the other two correspond to a free scalar field. We therefore write
\begin{equation}
\hat{A}_i = A_i, \quad \hat{Z}_{i_1 \dots i_{d-2}z} = Z_{i_1 \dots i_{d-2}}, \quad \hat{A}_z = \varphi, \quad \hat{Z}_{i_1 \dots i_{d-1}} = (-1)^d (d-1)!\, \omega_{i_1 \dots i_{d-1}} .
\end{equation}
The magnetic fields of $\hat{A}$ and $\hat{Z}$ then reduce as
\begin{equation}
\cB_{i_1 \dots i_{d-1}}[\hat{A}] = \varepsilon_{i_1 \dots i_{d-1} j} \partial^j \varphi, \quad \cB_{i_1 \dots i_{d-2} z}[\hat{A}] = \cB_{i_1 \dots i_{d-2}}[A]
\end{equation}
and
\begin{equation}
\cB_i[\hat{Z}] = \cB_i[Z], \quad \cB_z[\hat{Z}] = \varepsilon_{i j_1 \dots j_{d-1}} \partial^i \omega^{j_1 \dots j_{d-1}} = \pi[\omega]
\end{equation}
respectively. The action then becomes
\begin{align}
S[A,Z,\varphi,\omega] = \int \dtdx[d] \,\Big( &\cB_i[Z] \dot{A}^i - \frac{1}{2} \cB_i[Z] \cB^i[Z] - \frac{1}{2(d-2)!} \cB_{i_1 \dots i_{d-2}}[A]\cB^{i_1 \dots i_{d-2}}[A] \nonumber\\
& + \pi[\omega] \dot{\varphi}- \frac{1}{2} \pi[\omega]^2 - \frac{1}{2} \partial_i \varphi \, \partial^i \varphi \Big),
\end{align}
which is indeed the sum of a free vector action and a free scalar action. Remark that the momentum $\pi$ is written as the curl of a rank $(d-1)$ antisymmetric tensor in this reduction. This is always allowed and does not restrict $\pi$ (this is also discussed in \cite{Bunster:2011qp}).

\section{Dimensional reduction for the graviton and gravitino}
\label{app:dimred}

In this section, we perform the dimensional reduction on $S^1$ of the actions of sections \ref{sec:graviton} and \ref{sec:gravitino}, recovering familiar results in the prepotential formulation. We reduce from $D+1$ dimensions to $D$ dimensions along the last spatial coordinate, which we denote by $z$. Only the massless Kaluza-Klein mode is kept, i.e., all derivatives with respect to $z$ are set to zero.

Just like for the reduction of the vector field recalled above, a simple counting of derivatives shows that some of the lower-dimensional fields will be written as derivatives of a more basic quantity. It is indeed what we find and brings no loss of generality.

\subsection{Graviton}

\paragraph{Lagrangian formulation.}

We take the higher-dimensional graviton as
\begin{equation}\label{eq:ansatz}
\hat{h}_{\mu\nu} = h_{\mu\nu} + 2 \alpha \,\varphi \,\eta_{\mu\nu}, \quad \hat{h}_{\mu z} = A_\mu, \quad \hat{h}_{zz} = 2 \beta \,\varphi
\end{equation}
for some constants $\alpha$, $\beta$. This ansatz can be inverted to
\begin{equation}
h_{\mu\nu} = \hat{h}_{\mu\nu} - \frac{\alpha}{\beta} \hat{h}_{zz} \eta_{\mu\nu}, \quad A_{\mu} = \hat{h}_{\mu \,z}, \quad \varphi = \frac{1}{2\beta} \hat{h}_{zz} .
\end{equation}
Therefore, it is a good parametrization of the higher-dimensional metric as long as $\beta \neq 0$.

Then, the Lagrangian \eqref{LPF} reduces to the sum of the free Lagrangians for $h_{\mu\nu}$, $A_\mu$ and $\varphi$:
\begin{equation}\label{Lreduced}
\mathcal{L}^{(D+1)}_\text{PF}[\hat{h}] = \mathcal{L}^{(D)}_\text{PF}[h] - \frac{1}{4} F_{\mu\nu} F^{\mu\nu} - \frac{1}{2} \partial_\mu \varphi \partial^\mu \varphi ,
\end{equation}
provided the constants $\alpha$, $\beta$ are given by
\begin{equation}
\alpha^2 = \frac{1}{2(D-2)(D-1)}, \quad \beta = - \alpha (D-2).
\end{equation}
This is of course consistent with the usual KK ansatz at the non-linear level (see for example \cite{Cremmer:1997ct}).

\paragraph{Hamiltonian variables.}

From equations \eqref{eq:ansatz}, we get the reduction formulas
\begin{align}
\hat{h}_{ij} &= h_{ij} + 2 \alpha \,\varphi \,\delta_{ij}, &\hat{h}_{i z} &= A_i, &\hat{h}_{zz} &= 2 \beta \varphi, \\
\hat{\pi}^{ij} &= \pi^{ij}, &\hat{\pi}^{iz} &= \frac{1}{2} \pi^i, &\hat{\pi}^{zz} &= \frac{1}{2\beta}( \pi - 2 \alpha \, \delta_{ij} \pi^{ij}). 
\end{align}
with inverse relations
\begin{align}
h_{ij} &= \hat{h}_{ij} - \frac{\alpha}{\beta} \delta_{ij} \hat{h}_{zz}, &A_i &= \hat{h}_{i\,z}, &\varphi &= \frac{1}{2\beta} \hat{h}_{zz} ,\label{eq:ansatzh}\\
\pi^{ij} &= \hat{\pi}^{ij}, &\pi^i &= 2 \hat{\pi}^{iz}, &\pi &= 2 \beta \hat{\pi}^{zz} + 2 \alpha \delta_{ij} \hat{\pi}^{ij} .\label{eq:ansatzpi}
\end{align}
Plugging this in the higher-dimensional Hamiltonian action \eqref{eq:hamgraviton}, direct computation shows that it indeed reduces to the sum of the Hamiltonian actions \eqref{eq:Hs}, \eqref{eq:Hv} and \eqref{eq:hamgraviton} for a scalar field, a vector field and linearized gravity in $D=d+1$ dimensions.

\paragraph{Reduction of the first prepotential.}

The prepotential $\Phi_{I_1 \dots I_{d-1}J}$ splits into four pieces,
\begin{equation}\label{eq:phired}
\Phi_{i_1 \dots i_{d-1}j}, \quad \Phi_{z i_1 \dots i_{d-2} j}, \quad \Phi_{i_1 \dots i_{d-1} z}, \quad \Phi_{i_1 \dots i_{d-2} zz}.
\end{equation}
It is convenient to do the following field redefinition:
\begin{align}
Q_{i_1 \dots i_{d-1}j} &= \Phi_{i_1 \dots i_{d-1}j} - (d-1) \Phi_{[i_1 \dots i_{d-2} |zz|} \delta_{i_{d-1}]j}, \\
M_{i_1 \dots i_{d-2}j} &= \mathbb{P}_{(d-2,1)}(\Phi_{z i_1 \dots i_{d-2} j}) = \Phi_{z i_1 \dots i_{d-2} j} - \Phi_{z[i_1 \dots i_{d-2} j]},\\
A_{i_1 \dots i_{d-1}} &= \Phi_{z[i_1 \dots i_{d-1}]}, \\
B_{i_1 \dots i_{d-2}} &= \Phi_{i_1 \dots i_{d-2} zz} .
\end{align}
This change of variables is invertible: explicitely, one can express the quantities in \eqref{eq:phired} as
\begin{align}
\Phi_{i_1 \dots i_{d-1}j} &= Q_{i_1 \dots i_{d-1}j} + (d-1) B_{[i_1 \dots i_{d-2}} \delta_{i_{d-1}]j}, \\
\Phi_{z i_1 \dots i_{d-2} j} &= M_{i_1 \dots i_{d-2}j} + A_{i_1 \dots i_{d-2}j}, \label{eq:splitphiz}\\
\Phi_{i_1 \dots i_{d-1} z} &= (-1)^{d-1} (d-1) A_{i_1 \dots i_{d-1}}, \\
\Phi_{i_1 \dots i_{d-2} zz} &= B_{i_1 \dots i_{d-2}},
\end{align}
where we used the identity
\begin{equation}
\Phi_{z[i_1 \dots i_{d-2}j]} = \frac{(-1)^d}{d-1} \Phi_{i_1 \dots i_{d-2} j z},
\end{equation}
which follows from $\Phi_{[i_1 \dots i_{d-2}jz]}=0$.
The new variables have the irreducible Young symmetry suggested by their indices: $Q$ has the $(d-1,1)$ symmetry, $M$ the $(d-2,1)$ symmetry, and $A$ and $B$ are totally antisymmetric. Equation \eqref{eq:splitphiz} corresponds to the split of $\Phi_{z i_1 \dots i_{d-2} j}$ into its irreducible components.

The reduction of the gauge symmetries of the higher-dimensional prepotential are give the usual gauge symmetries for $Q$, $M$, $A$, $B$ according to their Young symmetry. For the reduction of the local Weyl transformations, one finds that
\begin{itemize}
\item $Q$ has no Weyl transformation (this motivates its definition);
\item $M$ has the same Weyl symmetries \eqref{eq:phigaugeapp} as the first prepotential of linearized gravity;
\item $A$ has no Weyl symmetry;
\item $B$ can be set to zero.
\end{itemize}
The link between these fields and the appropriate prepotentials for the lower-dimensional fields is done by using equations \eqref{eq:ansatzh}.
This gives
\begin{align}
h_{ij} &= 2(-1)^d(d-1) \,\varepsilon_{l_1 \dots l_{d-2} k(i} \pd^k M\indices{^{i_1 \dots i_{d-2}}_{j)}}, \\
A_i &= \varepsilon_{j_1 \dots j_{d-1} k} \pd^k Q\indices{^{j_1 \dots j_{d-1}}_i}, \\
\varphi &= \frac{1}{2\beta} (-1)^{d-1} (d-1) \,\varepsilon_{j_1 \dots j_{d-1}j} \pd^j A^{i_1 \dots i_{d-1}} .
\end{align}
Therefore, we identify the lower-dimensional prepotential $\phi$ as
\begin{equation}
\phi\indices{^{i_1 \dots i_{d-2}}_{j}} = (-1)^d(d-1) \, M\indices{^{i_1 \dots i_{d-2}}_{j}} .
\end{equation}

\paragraph{Reduction of the second prepotential.}

The prepotential $P_{I_1 \dots I_{d-1}J_1 \dots J_{d-1}}$ splits into three pieces,
\begin{equation}\label{eq:pred}
P_{i_1 \dots i_{d-1}j_1 \dots j_{d-1}}, \quad P_{i_1 \dots i_{d-1}j_1 \dots j_{d-2} z}, \quad P_{i_1 \dots i_{d-2} z j_1 \dots j_{d-2} z}.
\end{equation}
As before, it is useful to introduce a combination that has no Weyl symmetry,
\begin{equation}
T\indices{^{m_1 \dots m_{d-1}}_{n_1 \dots n_{d-1}}} = P\indices{^{m_1 \dots m_{d-1}}_{n_1 \dots n_{d-1}}} - (d-1) \delta^{[m_1}_{[n_1} P\indices{^{m_2 \dots m_{d-1}]z}_{n_2 \dots n_{d-1}] z}} .
\end{equation}
The lower-dimensional momenta are then
\begin{align}
\pi^{ij} &= G_{ij}[P] \nn \\
&= (d-1)^2 \varepsilon^{ikm_1 \dots m_{d-2}}\varepsilon^{jln_1 \dots n_{d-2}} \pd_k \pd_l P_{m_1 \dots m_{d-2}zn_1 \dots n_{d-2}z} \\
\pi^i &= 2 G^{iz}[P] \nn \\
&= \frac{1}{(d-2)!}\varepsilon^{ikm_1 \dots m_{d-2}} \pd_k \left( 2(-1)^d(d-1)!\, \varepsilon^{ln_1 \dots n_{d-1}} \pd^l P\indices{^{n_1 \dots n_{d-1}}_{ m_1 \dots m_{d-2} z}} \right) \\
\pi &= 2 \beta G^{zz}[P] + 2 \alpha \delta_{ij} G^{ij}[P] \nn \\
&= 2 \beta \varepsilon^{k m_1 \dots m_{d-1}}\varepsilon^{j n_1 \dots n_{d-1}} \pd_k \pd_l T_{m_1 \dots m_{d-1} n_1 \dots n_{d-1}} .
\end{align}
And we identify the potentials as
\begin{align}
P_{m_1 \dots m_{d-2}n_1 \dots n_{d-2}} &= (d-1)^2 P_{m_1 \dots m_{d-2}zn_1 \dots n_{d-2}z} , \\
Z_{m_1 \dots m_{d-2}} &= 2(-1)^d(d-1)!\, \varepsilon^{ln_1 \dots n_{d-1}} \pd^l P\indices{^{n_1 \dots n_{d-1}}_{ m_1 \dots m_{d-2} z}}.
\end{align}

\subsection{Gravitino}

\subsubsection{Covariant form}

It is useful to first write the reduction in covariant form. We have the three terms
\begin{align}
-\bar{\Psi}_{\hat{\mu}} \hat{\gamma}^{\hat{\mu}\hat{\nu}\hat{\rho}} \partial_{\hat{\nu}} \Psi_{\hat{\rho}} &= - \bar{\Psi}_z \hat{\gamma}^z \hat{\gamma}^{\mu\nu} \partial_\mu \Psi_\nu - \bar{\Psi}_\mu \hat{\gamma}^{\mu\nu} \hat{\gamma}^z \partial_\nu \Psi_z - \bar{\Psi}_\mu \hat{\gamma}^{\mu\nu\rho} \partial_\nu \Psi_\rho \nonumber \\
&= \bar{\eta} \hat{\gamma}^{\mu\nu} \partial_\mu \Psi_\nu - \bar{\Psi}_\mu \hat{\gamma}^{\mu\nu} \partial_\nu \eta - \bar{\Psi}_\mu \hat{\gamma}^{\mu\nu\rho} \partial_\nu \Psi_\rho,
\end{align}
where in the second line we defined $\eta = \hat{\gamma}^z \Psi_z$. (This definition implies $\bar{\eta} = - \bar{\Psi}_z \hat{\gamma}^z$.)
We now take the ansatz
\begin{equation}
\eta = a \lambda, \quad \Psi_\mu = \psi_\mu + b \hat{\gamma}_\mu \lambda,
\end{equation}
where $a$ and $b$ are constants. Requiring that the cross-terms between $\lambda$ and $\psi$ vanish first gives $a=-b(D-2)$. The constant $b$ then gives the normalization of the part containing $\lambda$: we choose $b^2(D-1)(D-2)=1$ for convenience. This gives the constants $a$, $b$ as
\begin{equation}
a = -\sqrt{\frac{D-2}{D-1}}, \qquad b = \frac{1}{\sqrt{(D-1)(D-2)}} .
\end{equation}
The action for the gravitino then reduces to
\begin{equation}\label{eq:actiongravitinored}
S[\lambda, \psi] = - \int \dx \left(  \bar{\lambda} \hat{\gamma}^\mu \partial_\mu \lambda + \bar{\psi}_\mu \hat{\gamma}^{\mu\nu\rho} \partial_\nu \psi_\rho \right).
\end{equation}
The ansatz above can be inverted to
\begin{align}\label{eq:gravitinored}
\lambda = - \sqrt{\frac{D-1}{D-2}}\,\hat{\gamma}^z \Psi_z, \quad \psi_\mu = \Psi_\mu + \frac{1}{D-2}\, \hat{\gamma}_\mu \hat{\gamma}^z \Psi_z .
\end{align}
Note that the action \eqref{eq:actiongravitinored} is not quite the action for a free Dirac field and a free gravitino in $D$ dimensions, since it still contains the higher-dimensional gamma matrices $\hat{\gamma}_\mu$. To go further, we must specify the link between gamma matrices in $D+1$ and $D$ dimensions.

\subsubsection{Choice of gamma matrices.}

We now set up our choice of gamma matrix representations. Starting from a given representation (that we do not need to specify) in even dimension $D = 2m$, we go up in dimensions and give representations in dimensions $2m+1$ and $2m+2$.

In odd dimensions $D = 2m +1$, we can take the first $2m$ matrices to be exactly those in dimension $D= 2m$, and the last one to be $\gamma_*$. Indeed, because of the first two equations of \eqref{eq:gammastarprop}, the set
\begin{equation}
(\hat{\gamma}_\mu) = ( \gamma_0, \gamma_1, \dots, \gamma_{2m-1}, \gamma_{2m} = \gamma_* )
\end{equation}
satisfies the defining property \eqref{eq:cliff} in dimension $2m + 1$. (Note that we could also take $\gamma_{2m} = - \gamma_*$: this gives an equivalent representation.)

In dimension $D = 2m + 2$, one can take the gamma matrices in the block form
\begin{align}\label{eq:explicitgamma}
\hat{\gamma}_\mu &= \sigma_1 \otimes \gamma_\mu = \begin{pmatrix}
0 & \gamma_\mu \\
\gamma_\mu & 0
\end{pmatrix} \quad(\mu = 0, \dots, D-2), \\
\hat{\gamma}_{D-1} &= \sigma_2 \otimes I= \begin{pmatrix}
0 & - iI \\ iI & 0
\end{pmatrix}, \nonumber
\end{align}
The first are given in terms of the $D-1 = 2m + 1$ gamma matrices (in particular, $\gamma_{2m} = (-i)^{m+1} \gamma_0 \gamma_1 \dots \gamma_{2m-1}$ is the chirality matrix in $D-2 = 2m$ dimensions). In this representation, the $\hat{\gamma}_*$ matrix is diagonal,
\begin{equation}
\hat{\gamma}_* = \begin{pmatrix}
I & 0 \\ 0 & -I
\end{pmatrix} .
\end{equation}

\subsubsection{Odd to even dimension.}

\paragraph{Covariant formulation.}

The gamma matrices in $D+1$ and $D$ dimensions have the same size, and so do the spinors. The lower-dimensional fields are, using $\hat{\gamma}_\mu = \gamma_\mu$ and $\hat{\gamma}_z = \gamma_*$,
\begin{align}\label{eq:gravitinoredodd}
\lambda = - \sqrt{\frac{D-1}{D-2}}\,\gamma_* \Psi_z, \quad \psi_\mu = \Psi_\mu + \frac{1}{D-2}\, \gamma_\mu \gamma_* \Psi_z .
\end{align}
The action \eqref{eq:actiongravitinored} is just the sum of the free action for a Dirac field $\lambda$ and a Rarita-Schwinger field $\psi_\mu$ in $D$ dimensions. Note also that redefinitions of the fields by a phase may be required to accommodate for (Majorana or symplectic-Majorana) reality conditions in $D$ dimensions; this poses no difficulty and will not be done here.

\paragraph{Prepotential formulation.}

The prepotential reduces to two pieces,
\begin{equation}
X_{I_1 \dots I_{d-1}} \;\rightarrow\; X_{i_1 \dots i_{d-1}},\; X_{i_1 \dots i_{d-2} z} ,
\end{equation}
with the associated gauge and Weyl transformations
\begin{align}
\delta X_{i_1 \dots i_{d-1}} &= (d-1) \partial_{[i_1} \Lambda_{i_2 \dots i_{d-1}]} + \gamma_{i_1 \dots i_{d-1}} W \\
\delta X_{i_1 \dots i_{d-2} z} &= (d-2) \partial_{[i_1} \Lambda_{i_2 \dots i_{d-2}] z} + \gamma_{i_1 \dots i_{d-2}} \gamma_* W \label{eq:chigaugered}.
\end{align}
To identify those prepotentials with those of the $D$-dimensional fields, one must compare with \eqref{eq:gravitinoredodd}. For that, one first needs the reduction of the Schouten tensor of $X$ (indeed, the higher-dimensional gravitino is determined by $X$ through the equation $\Psi_I = S_I[X]$). One finds
\begin{align}
S_i[X] &= \frac{1}{d} \left( \gamma_{ij} - (d-1) \delta_{ij} \right) G^j[X] + \gamma_i \gamma_* G_z[X], \\
S_z[X] &= - \frac{1}{d} \left( \gamma_i \gamma_* G^i[X] + (d-1) G_z[X] \right) .
\end{align}
in terms of the Einstein tensor. The lower-dimensional fields are then
\begin{align}
\psi_i &= S_i[X] + \frac{1}{d-1} \gamma_i \gamma_* S_z[X] = \frac{1}{d-1} \left( \gamma_{ij} - (d-2) \delta_{ij} \right) G^j[X] \label{eq:psiofX} \\
\lambda &= - \sqrt{\frac{d}{d-1}} \gamma_* S_z[X] = \frac{1}{\sqrt{d(d-1)}} \left( -\gamma_i G^i[X] + (d-1) \gamma_* G_z[X] \right) .
\end{align}
The Einstein itself reduces as
\begin{equation}
G_i[X] = (d-1) \varepsilon_{ij k_1 \dots k_{d-2}} \partial^j X^{k_1 \dots k_{d-2} z}, \quad G_z[X] = \varepsilon_{i j_1 \dots j_{d-1}} \partial^i X^{j_1 \dots j_{d-1}}.
\end{equation}
Therefore, \eqref{eq:psiofX} is exactly the formula $\psi_i = S_i[\chi]$ relating the lower-dimensional gravitino to its prepotential, that we identify as
\begin{equation}
\chi_{i_1 \dots i_{d-2}} = (d-1) X_{i_1 \dots i_{d-2} z}
\end{equation}
(the gauge and Weyl symmetries also match, see \eqref{eq:chigaugered}). For the Dirac field $\lambda$, one finds
\begin{equation}
\lambda = \varepsilon_{i_1 \dots i_d} \partial^{i_1} \zeta^{i_2 \dots i_d}, \quad \zeta_{i_1 \dots i_{d-1}} = \sqrt{\frac{d-1}{d}} \left( \gamma_* X_{i_1 \dots i_{d-1}} + \gamma_{[i_1} X_{i_2 \dots i_{d-1}]z} \right) .
\end{equation}
The gauge transformation of the right-hand side is easily checked to give $\delta \lambda = 0$, as it should. Note also that the change of variables  from the two $X$'s to the pair $(\chi, \zeta)$ is invertible.

This identification of the fields now implies the reduction of the action \eqref{eq:actionchi} as above.

\subsubsection{Even to odd dimension.}

\paragraph{Covariant formulation.}

This case is a bit more involved, since the gamma matrices in even dimension $D+1$ now have twice the size of those in one dimension below. Accordingly, we write the higher-dimensional gravitino in the block form
\begin{equation}
\Psi_{\mu} = \col{\Psi^+_\mu}{\Psi^-_\mu} .
\end{equation}
Note that a chirality condition $\Gamma_* \Psi_\mu = \pm \Psi_\mu$ on the higher-dimensional gravitino corresponds to keeping only the component $\Psi^\pm$ in this decomposition.

According to \eqref{eq:gravitinored}, the $D$-dimensional fields are then
\begin{equation}\label{eq:psipm}
\psi_\mu = \col{\psi^+_\mu}{\psi^-_\mu}, \quad \psi^\pm_\mu = \Psi^\pm_\mu \pm \frac{i}{D-2} \, \gamma_\mu \Psi^\pm_z
\end{equation}
and
\begin{equation}\label{eq:lambdapm}
\lambda = \col{\lambda^+}{\lambda^-}, \quad \lambda^\pm = \pm i \sqrt{\frac{D-1}{D-2}} \Psi^\mp_z.
\end{equation}
The action \eqref{eq:actiongravitinored} then reduces to the sum of two Dirac and two gravitino actions,
\begin{equation}
S[\lambda^\pm, \psi^\pm] = - \int \dx \left( \bar{\lambda}^+ \gamma^\mu \partial_\mu \lambda^+ + \bar{\lambda}^- \gamma^\mu \partial_\mu \lambda^- + \bar{\psi}^+_\mu \gamma^{\mu\nu\rho} \partial_\nu \psi^+_\rho + \bar{\psi}^-_\mu \gamma^{\mu\nu\rho} \partial_\nu \psi^-_\rho \right) .
\end{equation}
Again, phases can be incorporated in the various fields to accommodate for reality conditions in $D$ dimensions.

\paragraph{Prepotential formulation.}

We write the prepotential for $\Psi_\mu$ in the block form
\begin{equation}\label{eq:chiblock}
X_{I_1 \dots I_{d-1}} = \col{X^+_{I_1 \dots I_{d-1}}}{X^-_{I_1 \dots I_{d-1}}} .
\end{equation}
One has therefore four fields after reduction,
\begin{equation}
X^\pm_{I_1 \dots I_{d-1}} \;\rightarrow\; X^\pm_{i_1 \dots i_{d-1}},\; X^\pm_{i_1 \dots i_{d-2} z} ,
\end{equation}
with the associated gauge and Weyl transformations
\begin{align}
\delta X^\pm_{i_1 \dots i_{d-1}} &= (d-1) \partial_{[i_1} \Lambda^\pm_{i_2 \dots i_{d-1}]} + \gamma_{i_1 \dots i_{d-1}} W^\mp \\
\delta X^\pm_{i_1 \dots i_{d-2} z} &= (d-2) \partial_{[i_1} \Lambda^\pm_{i_2 \dots i_{d-2}] z} \mp i \gamma_{i_1 \dots i_{d-2}} W^\mp \label{eq:psigaugereduced},
\end{align}
where $\Lambda^\pm$ and $W^\pm$ are defined from the higher-dimensional parameters by a block form analogous to \eqref{eq:chiblock}.

To identify those prepotentials with those of $D$-dimensional fields, one must compare with equations \eqref{eq:psipm} and \eqref{eq:lambdapm}. For that, one first needs the reduction of the Schouten tensor of $X$ (indeed, the higher-dimensional gravitino is determined by $X$ through the equation $\Psi_I = S_I[X]$). One finds
\begin{align}
S_i^\pm[X] &= \frac{1}{d} \left( \gamma_{ij} - (d-1) \delta_{ij} \right) G^{\pm j}[X] \pm \frac{i}{d} \gamma_i G^\pm_z[X], \\
S^\pm_z[X] &= \frac{1}{d} \left( \mp i \gamma_i G^{\pm i}[X] - (d-1) G^\pm_z[X] \right) .
\end{align}
in terms of the Einstein tensor. The lower-dimensional fields are therefore
\begin{align}
\psi_i^\pm &= S^\pm_i[X] \pm \frac{i}{d-1} \gamma_i S^\pm_z[X] = \frac{1}{d-1} \left( \gamma_{ij} - (d-2) \delta_{ij} \right) G^{\pm j}[X] \label{eq:psipmofX} \\
\lambda^\pm &= \pm i \sqrt{\frac{d}{d-1}} S^\mp_z[X] = \frac{1}{\sqrt{d(d-1)}} \left( - \gamma_i G^{\mp i}[X] \mp i (d-1) G^\mp_z[X] \right) .
\end{align}
The Einstein itself reduces as
\begin{equation}
G^\pm_i[X] = (d-1) \varepsilon_{ij k_1 \dots k_{d-2}} \partial^j X^{\pm k_1 \dots k_{d-2} z}, \quad G^\pm_z[X] = \varepsilon_{i j_1 \dots j_{d-1}} \partial^i X^{\pm j_1 \dots j_{d-1}}.
\end{equation}
Therefore, \eqref{eq:psipmofX} is exactly the formula $\psi^\pm_i = S_i[\chi^\pm]$ relating the lower-dimensional gravitinos to their prepotentials
\begin{equation}
\chi^\pm_{i_1 \dots i_{d-2}} = (d-1) X^\pm_{i_1 \dots i_{d-2} z}
\end{equation}
(the gauge and Weyl symmetries also match with the appropriate identifications, see \eqref{eq:psigaugereduced}). For the Dirac fields $\lambda^\pm$, one finds
\begin{equation}
\lambda^\pm = \varepsilon_{i_1 \dots i_d} \partial^{i_1} \zeta^{\pm i_2 \dots i_d}, \quad \zeta^{\pm}_{i_1 \dots i_{d-1}} = \sqrt{\frac{d-1}{d}} \left( \mp i X^\mp_{i_1 \dots i_{d-1}} + \gamma_{[i_1} X^\mp_{i_2 \dots i_{d-1}]z} \right) .
\end{equation}

\section{Dimensional reduction of self-dual fields in six dimensions}
\label{app:40}

In this appendix, we derive the dimensional reduction of some six-dimensional fields whose curvature satisfies a self-duality condition. Those fields appear in the intriguing $\mathcal{N} = (4,0)$ maximally supersymmetric theory in six dimensions \cite{Hull:2000zn,Hull:2000rr}. Their six-dimensional actions were written with prepotentials in references \cite{Henneaux:2016opm,Henneaux:2017xsb} and directly reduce to the actions considered in this paper for linearized supergravity.

\subsection{The exotic graviton}

The exotic graviton is a bosonic field $T_{\mu\nu\rho\sigma}$ with the $(2,2)$ symmetry
\begin{equation}
T_{\mu\nu\rho\sigma} = T_{[\mu\nu]\rho\sigma} = T_{\mu\nu[\rho\sigma]}, \quad T_{[\mu\nu\rho]\sigma} = 0 .
\end{equation}
Its curvature tensor
\begin{equation}
R\indices{^{\mu\nu\rho}_{\alpha\beta\gamma}}[T] = \partial^{[\mu}\partial_{[\alpha} T\indices{^{\nu\rho]}_{\beta\gamma]}}
\end{equation}
satisfies the self-duality equation
\begin{equation}\label{eq:6dexgrav}
R_{\alpha_1 \alpha_2 \alpha_3 \beta_1 \beta_2 \beta_3}[T] = \frac{1}{3!} \varepsilon_{\alpha_1 \alpha_2 \alpha_3 \lambda_1 \lambda_2 \lambda_3} R\indices{^{\lambda_1 \lambda_2 \lambda_3}_{\beta_1 \beta_2 \beta_3} }[T].
\end{equation}
The gauge invariances of $T$ are $\delta T\indices{^{\mu\nu}_{\rho\sigma}} = \partial^{[\mu} \alpha\indices{_{\rho\sigma}^{\nu]}} + \partial_{[\rho} \alpha\indices{^{\mu\nu}_{\sigma]}}$, where the gauge parameter has the $(2,1)$ symmetry.

\paragraph{Prepotential formulation.}

The action yielding the equations of motion \eqref{eq:6dexgrav} was first written in \cite{Henneaux:2016opm}, and it involves prepotentials in an essential way. The prepotential is a $(2,2)$ tensor $Z_{IJKL}$, determined up to the local gauge and Weyl transformations
\begin{equation} \label{eq:6d22gauge}
\delta Z_{IJKL} = \xi_{IJ[K,L]} + \xi_{KL[I,J]} + \delta_{[I[K} \lambda_{J]L]},
\end{equation}
where $\xi_{IJK}$ is a $(2,1)$ tensor parametrizing the gauge transformations and $\lambda_{IJ}$ is a symmetric tensor parametrizing the Weyl rescalings.

The Einstein, Schouten and Cotton tensors of $Z$ are
\begin{align}
G\indices{^{IJ}_{KL}}[Z] &= \frac{1}{3!^2} \varepsilon^{IJABC} \varepsilon_{KLPQR} \partial_A \partial^P Z\indices{_{AB}^{QR}}, \\
S\indices{^{IJ}_{KL}}[Z] &= G\indices{^{IJ}_{KL}}[Z] - 2 \delta^{[I}_{[K} G\indices{^{J]}_{L]}}[Z] + \frac{1}{3} \delta^I_{[K} \delta^J_{L]} G[Z], \\
D_{IJKL}[Z] &= \frac{1}{3!} \varepsilon_{IJABC} \partial^{A} S\indices{^{BC}_{KL}}[Z],
\end{align}
where the traces of the Einstein tensor are defined by $G\indices{^I_J} = G\indices{^{IK}_{JK}}$ and $G = G\indices{^{IJ}_{IJ}}$.
The action is
\begin{equation}\label{eq:6d22action}
S[Z] = \frac{1}{2} \int \! d^6\!x \, Z_{MNRS} \left( \dot{D}^{MNRS}[Z] - \frac{1}{2} \varepsilon^{MNIJK} \partial_K D\indices{_{IJ}^{RS}}[Z] \right) .
\end{equation}

\paragraph{Reduction of the field and gauge transformations.}

The prepotential $Z_{IJKL}$ and its gauge parameters split into several pieces,
\begin{align}
Z_{IJKL} \;&\longrightarrow\; Z_{ijkl},\; Z_{ij k5},\; Z_{i5j5}, \\
\xi_{IJK} \;&\longrightarrow\; \xi_{ijk},\; \xi_{i5j},\; \xi_{ij5},\; \xi_{i55}, \\
\lambda_{IJ} \;&\longrightarrow\; \lambda_{ij},\; \lambda_{i5},\; \lambda_{55}.
\end{align}
The transformation of $Z_{i5j5}$ under Weyl rescalings is
\begin{equation}
\delta Z_{i5j5} = \frac{1}{4} ( \lambda_{ij} + \delta_{ij} \lambda_{55} ) .
\end{equation}
Therefore, one can gauge away $Z_{i5j5}$ by a Weyl transformation. We will set $Z_{i5j5} = 0$ from now on. To preserve the condition, one must then set $\xi_{i55} = 0$ and $\lambda_{ij} = - \delta_{ij} \lambda_{55}$. 
Also, we can split $\xi_{i5j} = - \xi_{5ij}$ into its symmetric and antisymmetric parts,
\begin{equation}
\xi_{5ij} = s_{ij} + a_{ij}, \qquad s_{ij} = \xi_{5(ij)}, \qquad a_{ij} = \xi_{5[ij]} .
\end{equation}
The cyclic identity $3\,\xi_{[IJK]} = \xi_{IJK} + \xi_{JKI} + \xi_{KIJ} = 0$ implies that the $\xi_{ij5}$ component is not independent: $\xi_{ij5} = - 2 a_{ij}$.
The gauge transformations of the remaining fields are then
\begin{align}
\delta Z_{ijkl} &= \xi_{ij[k,l]} + \xi_{kl[i,j]} - \frac{1}{2} (\delta_{ik} \delta_{jl} - \delta_{il} \delta_{jk} ) \lambda_{55} \\
\delta Z_{ijk5} &= \partial_k a_{ij} + \partial_{[i} s_{j]k} - \partial_{[i} a_{j]k} + \frac{1}{2} \delta_{k[i} \lambda_{j]5}. 
\end{align}
Those are exactly the gauge transformations \eqref{eq:phigaugeapp} and \eqref{eq:Pgaugeapp} for the prepotentials of linearized gravity in five dimensions, provided we identify the fields and gauge parameters as
\begin{align}
P_{ijkl} &= Z_{ijkl}, \quad \phi_{ijk} = Z_{ijk5}, \\
\alpha_{ijk} &= \xi_{ijk},  \quad A_{ij} = a_{ij}, \quad M_{ij} = s_{ij}, \quad \xi = -\lambda_{55}, \quad B_i = \frac{1}{2}\lambda_{i5} .
\end{align}

\paragraph{Reduction of the curvature tensors.} The Einstein, Schouten and Cotton tensors of $Z_{IJKL}$ reduce as follows:
\begin{itemize}
\item Einstein:
\begin{align}
G_{ijkl}[Z] = 0, \quad G_{ijk5}[Z] = - \frac{1}{18} G_{ijk}[\phi], \quad G_{i5j5}[Z] = \frac{1}{3!^2} G_{ij}[P],
\end{align}
\item Schouten:
\begin{align}
S_{ijkl}[Z] = - \frac{1}{18} \delta_{[i[k} S_{l]j]}[P], \quad S_{ijk5}[Z] = - \frac{1}{18} S_{ijk}[\phi], \quad S_{i5j5}[Z] = \frac{1}{2.3!^2} S_{ij}[P],
\end{align}
\item Cotton:
\begin{align}
D_{ijkl}[Z] = - \frac{1}{3.18} D_{ijkl}[\phi], \quad D_{ijk5}[Z] = \frac{1}{3!^3} D_{ijk}[P], \quad D_{i5j5}[Z] = \frac{1}{3.18} D\indices{_i^k_{jk}}[\phi],
\end{align}
\end{itemize}
where the right-hand sides are given by the corresponding tensors of sections \ref{app:confphi} and \ref{app:confP} for the graviton prepotentials.

\paragraph{Reduction of the action.} We can now use those formulas to reduce the six-dimensional action \eqref{eq:6d22action}. This gives
\begin{align}
S[\phi,P] = &\frac{1}{3.36} \int \dtdx[5] \left( \phi_{ijk} \dot{D}^{ijk}[P] - P_{ijkl} \dot{D}^{ijkl}[\phi] \right) \\
&-\frac{1}{9.6} \int \dtdx[5] \left( \phi_{ijk} \varepsilon^{kabc} \partial_a D\indices{_{bc}^{ij}} + \frac{1}{8}P_{ijkl} \varepsilon^{ijab}[\phi] \partial_a D\indices{^{kl}_b}[P] \right) . \nonumber
\end{align}
This is the action of section \ref{sec:actiongraviton}, up to the redefinitions $P_{ijkl} = 12 \sqrt{3} \,P'_{ijkl}$ and $\phi_{ijk} = 3 \sqrt{3} \,\phi'_{ijk}$. This result was annoucend in \cite{Henneaux:2016opm}\footnote{The sign discrepancy with respect to the appendix C of that reference comes from the fact that the prepotential $\phi_{ijk}$ we use here differs by a sign from the one of reference \cite{Bunster:2013oaa} (see also footnote \ref{footnote:sign} on page \pageref{footnote:sign}).} but is made more transparent by using the appropriate Cotton tensors in five dimensions.

\subsection{The exotic gravitino}

The exotic gravitino is a left-handed fermionic $2$-form $\Psi_{\mu\nu}$,
\begin{equation}
\Psi_{\mu\nu} = \Psi_{[\mu\nu]}, \quad \Gamma_* \Psi_{\mu\nu} = \Psi_{\mu\nu} .
\end{equation}
It satisfies the generalized Rarita-Schwinger equation
\begin{equation}\label{eq:eom2form}
\Gamma^{\mu\nu\alpha\beta\gamma} H_{\alpha\beta\gamma} = 0,
\end{equation}
where
$H_{\mu\nu\rho} = 3 \partial_{[\mu} \Psi_{\nu\rho]}$
is the field strength of $\Psi_{\mu\nu}$. It is invariant under the gauge transformation $\delta \Psi_{\mu\nu} = 2 \partial_{[\mu} \lambda_{\nu]}$ .

\paragraph{Prepotential formulation.} 
In reference \cite{Henneaux:2017xsb}, it was shown that the equation of motion \eqref{eq:eom2form} is equivalent to the self-duality equation $H = \star H$ on the field strength, i.e.
\begin{equation}
H_{\mu\nu\rho} = \frac{1}{3!} \varepsilon_{\mu\nu\rho\sigma\tau\lambda} H^{\sigma\tau\lambda},
\end{equation}
supplemented by the purely spatial constraint
\begin{equation}
\Gamma^{IABC} H_{ABC} = 0.
\end{equation}
Again, this constraint can be solved by the use of prepotentials; one finds that the appropriate prepotential is an antisymmetric left-handed spinor $X_{IJ}$, i.e.
\begin{equation}
X_{IJ} = X_{[IJ]}, \quad \Gamma_* X_{IJ} = X_{IJ} ,
\end{equation}
just like the spatial components $\Psi_{IJ}$ themselves. It is determined by $\Psi_{IJ}$ up to a gauge and Weyl transformation
\begin{equation}\label{eq:GaugeChi}
\delta X_{IJ} = \partial_{[I} \Lambda_{J]} + \Gamma_{[I} W_{J]} ,
\end{equation}
where the gauge parameters satisfy $\Gamma_* \Lambda_I = \Lambda_I$ and $\Gamma_* W_I = - W_I$ in order to preserve the chirality condition $\Gamma_* X_{IJ} = X_{IJ}$. The Einstein, Schouten and Cotton tensors of $X$ are given by
\begin{align}
G_{IJ}[X] &= \varepsilon_{IJKLM} \partial^K X^{LM}, \\
S_{IJ}[X] &= - \left( \delta^{[K}_{[I} \Gamma\indices{_{J]}^{L]}} + \frac{1}{6} \Gamma_{IJ} \Gamma^{KL} \right) G_{KL}[X], \\
D_{IJ}[X] &= \varepsilon_{IJKLM} \partial^K S^{LM}[X].
\end{align}
The spatial components $\Psi_{IJ}$ are given in terms of the prepotential by $\Psi_{IJ} = S_{IJ}[X]$. The action for $\Psi_{\mu\nu}$ can then be written as
\begin{align}\label{eq:exoticgravitino}
S[X] = -2i \int \!dt \,d^5\!x\, X^\dagger_{IJ} \left( \dot{D}^{IJ}[X] - \frac{1}{2} \varepsilon^{IJKLM} \partial_K D_{LM}[X] \right).
\end{align}

\paragraph{Reduction of the field and gauge transformations.}

For the dimensional reduction, we use the form \eqref{eq:explicitgamma} of the six-dimensional gamma matrices. In particular, the block-diagonal form of $\Gamma_*$ implies that the prepotential $X_{IJ}$ only has components in the first block,
\begin{equation}
X_{IJ} = \begin{pmatrix}
\hat{\chi}_{IJ} \\ 0
\end{pmatrix}.
\end{equation}
The field $\hat{\chi}_{IJ}$ then splits into two parts, $\hat{\chi}_{ij}$ and $\hat{\chi}_{i5}$. From \eqref{eq:GaugeChi}, we find their gauge transformations
\begin{align}
\delta \hat{\chi}_{ij} = \partial_{[i} \hat{\eta}_{j]} + \gamma_{[i} \hat{\rho}_{j]}, \quad \delta \hat{\chi}_{i5} = \frac{1}{2} (\partial_i \hat{\eta}_5 + \gamma_i \hat{\rho}_5 + i \hat{\rho}_i ),
\end{align}
where $\hat{\eta}_I$ and $\hat{\rho}_I$ are given from the six-dimensional gauge parameters by the block form
\begin{equation}
\Lambda_{I} = \begin{pmatrix}
\hat{\eta}_{I} \\ 0
\end{pmatrix},\qquad W_{I} = \begin{pmatrix}
0\\ \hat{\rho}_{I}
\end{pmatrix}.
\end{equation}
Using $\hat{\rho}_i$, the field $\hat{\chi}_{i5}$ can be set to zero. The residual gauge transformations must satisfy $\delta \hat{\chi}_{i5} = 0$ to respect this choice; this imposes $\hat{\rho}_i = i (\partial_i \hat{\eta}_5 + \gamma_i \hat{\rho}_5)$.
The gauge transformations of $\hat{\chi}_{ij}$ are then
\begin{equation}
\delta \hat{\chi}_{ij} = 2 \partial_{[i} \eta_{j]} + \gamma_{ij} \rho, \qquad \text{with} \quad \eta_j = \frac{1}{2} \left( \hat{\eta}_j - i \gamma_j \hat{\eta}_5\right), \quad \rho = i \hat{\rho}_5 .
\end{equation}
They have exactly the form \eqref{eq:chigaugeapp} of the gauge transformations of the prepotential for the five-dimensional gravitino.

\paragraph{Reduction of the curvature tensors.}

The various curvature tensors of $X_{IJ}$ reduce as follows:
\begin{itemize}
\item Einstein:
\begin{equation}
\hat{G}_{ij} = 0, \quad \hat{G}_{i5} = - G_i[\hat{\chi}] ,
\end{equation}
\item Schouten:
\begin{equation}
\hat{S}_{ij} = i \gamma_{[i} S_{j]}[\hat{\chi}], \quad \hat{S}_{i5} = \frac{1}{2} S_i[\hat{\chi}] ,
\end{equation}
\item Cotton:
\begin{equation}
\hat{D}_{ij} = D_{ij}[\hat{\chi}], \quad \hat{D}_{i5} = i \gamma^k D_{ik}[\hat{\chi}] ,
\end{equation}
\end{itemize}
where the left-hand sides are defined by the block forms
\begin{equation}
G_{IJ}[X] = \begin{pmatrix}
\hat{G}_{IJ} \\ 0
\end{pmatrix}, \quad
S_{IJ}[X] = \begin{pmatrix}
\hat{S}_{IJ} \\ 0
\end{pmatrix}, \quad
D_{IJ}[X] = \begin{pmatrix}
\hat{D}_{IJ} \\ 0
\end{pmatrix},
\end{equation}
and the right-hand sides are given by the appropriate tensors of section \ref{app:conffermion}.

\paragraph{Reduction of the action.} Using the above formulas, the action \eqref{eq:exoticgravitino} reduces to
\begin{align}
S[\chi] = -2i \int \!dt \,d^4\!x\, \hat{\chi}^\dagger_{ij} \left( \dot{D}^{ij}[\hat{\chi}] - i \varepsilon^{ijkl} \gamma^m \partial_k D_{lm}[\hat{\chi}] \right),
\end{align}
which is exactly the action \eqref{eq:actionchi} for the gravitino in five dimensions in the prepotential formalism, up to the redefinition $\hat{\chi}_{ij} = \chi_{ij} / \sqrt{2}$. This result was announced in \cite{Henneaux:2017xsb}.

\paragraph{Supersymmetry transformations.} The sum of actions \eqref{eq:6d22action} and \eqref{eq:exoticgravitino} is invariant under a supersymmetry transformation mixing the exotic graviton and gravitino, written explicitely in \cite{Henneaux:2017xsb}. Direct reduction of that variation does not reproduce the supersymmetry variations of section \ref{sec:susy}: this is because the formula written in \cite{Henneaux:2017xsb} does not preserve the gauge condition $\hat{\chi}_{i5} = 0$ in five dimensions. However, a pure gauge term may be added to the variation $\delta X_{IJ}$ in six dimensions to ensure $\delta \hat{\chi}_{i5} = 0$ upon reduction to five dimensions. Once this is done, we recover correctly the formulas of this paper for the supersymmetry variations of the prepotentials in five dimensions\footnote{This procedure also allows us to fix the overall constant $\alpha_1$ that appears in \cite{Henneaux:2017xsb} to $\alpha_1 = 3 \sqrt{6}$: this gives the usual factors of section \ref{sec:susy} for the variation of the five-dimensional metric and gravitino.}.

\subsection{The chiral two-form}

For completeness, we also review the reduction of the chiral two-form. The field is a two-form $\hat{A}_{\mu\nu} = \hat{A}_{[\mu\nu]}$, whose field strength $F_{\mu\nu\rho} = 3 \partial_{[\mu} \hat{A}_{\nu\rho]}$ satisfies the self-duality equation
\begin{equation}
F_{\mu\nu\rho} = \frac{1}{3!} \varepsilon_{\mu\nu\rho\alpha\beta\gamma} F^{\mu\nu\rho} .
\end{equation}
The field strength is invariant under the gauge transformations $\delta \hat{A}_{\mu\nu} = 2 \partial_{[\mu} \lambda_{\nu]}$ .

\paragraph{Quadratic action.}

The action for this field was first written in \cite{Henneaux:1988gg}; it is expressed in terms of the spatial components $\hat{A}_{IJ}$ only and reads
\begin{equation}
S[\hat{A}] = \frac{1}{2} \int \dtdx[5] \left( \dot{\hat{A}}_{IJ} \cB^{IJ}[\hat{A}] - \cB_{IJ}[\hat{A}] \cB^{IJ}[\hat{A}] \right),
\end{equation}
where $\cB^{IJ}[\hat{A}]$ is the magnetic field
\begin{equation}
\cB_{IJ}[\hat{A}] = \frac{1}{2} \varepsilon_{IJKLM} \partial^K \hat{A}^{LM} .
\end{equation}

\paragraph{Reduction of the field and gauge transformations.}

The field $\hat{A}_{IJ}$ reduces to two pieces,
\begin{equation}
\hat{A}_{IJ} \;\longrightarrow\; \hat{A}_{ij},\; \hat{A}_{i5} .
\end{equation}
Their gauge transformations are $\delta \hat{A}_{ij} = 2\partial_{[i} \lambda_{j]}$ and $\delta \hat{A}_{i5} = \partial_i \lambda_5$. Those are exactly the potentials and gauge transformations of the two-potential formulation of a free Maxwell field in five dimensions (see \cite{Bunster:2011qp} and appendix \ref{app:vector}). We therefore write
\begin{equation}
\hat{A}_{ij} = Z_{ij}, \quad \hat{A}_{i5} = A_i
\end{equation}
from now on.

\paragraph{Reduction of the magnetic field.}

The magnetic field splits as
\begin{equation}
\cB_{ij}[\hat{A}] = \cB_{ij}[A], \quad \cB_{i5}[\hat{A}] = - \cB_i[Z],
\end{equation}
where the right-hand sides are given by the five-dimensional magnetic fields defined in appendix \ref{app:vector}.

\paragraph{Reduction of the action.}

The action the becomes
\begin{equation}
S[A,Z] = \frac{1}{2} \int \dtdx[4] \left( \dot{Z}_{ij} \cB^{ij}[A] - 2 \dot{A}_i \cB^i[Z] - \cB_{ij}[A] \cB^{ij}[A] - 2 \cB_i[Z] \cB^i[Z] \right).
\end{equation}
Up to the rescalings $A_i = A'_i/\sqrt{2}$, $Z_{ij} = - Z'_{ij}/\sqrt{2}$, this is the two-potential action for a free vector field.

\section{Gamma matrices}
\label{app:gammamatrices}

We follow the conventions of \cite{VanProeyen:1999ni,freedman_vanproeyen_2012}. Gamma matrices are defined by
\begin{equation} \label{eq:cliff}
\{ \gamma^\mu, \gamma^\nu \} = 2 \eta^{\mu\nu},
\end{equation}
where the flat metric $\eta_{\mu\nu}$ is of ``mostly plus" signature, $\eta = \text{diag}(-+\dots  +)$. 
Useful identities on the spatial gamma matrices are
\begin{align}
\gamma_i \gamma_j &= \gamma_{ij} + \delta_{ij} \\
\gamma_i \gamma^{i j_1 \dots j_n} &= (d-n) \gamma^{j_1\dots j_n} \\
\frac{1}{d-1} \left( \gamma_{ij} - (d-2) \delta_{ij} \right) \gamma^{jk} &= \delta^k_i \label{eq:gammadelta} \\
\gamma_{ij} \gamma^{k_1 \dots k_n} &= \gamma\indices{_{ij}^{k_1 \dots k_n}} - 2 n\, \delta_{[i}^{[k_1} \gamma\indices{_{j]}^{k_2 \dots k_n]}} - n(n-1) \,\delta^{[k_1 k_2}_{ij} \gamma^{k_3 \dots k_n]} . \label{eq:g2gd-2}
\end{align}
(They are of course also valid when the indices are space-time indices, provided $\delta$ is repaced by $\eta$ and $d$ by $D$.)
We have the hermiticity properties\begin{equation} \label{gammaherm}
(\gamma^\mu)^\dagger = \gamma^0 \gamma^\mu \gamma^0 ,
\end{equation}
i.e. $(\gamma^0)^\dagger = - \gamma^0$ and $(\gamma^i)^\dagger = \gamma^i$ .
In even dimensions $D = 2m$, we can introduce the chirality matrix
\begin{equation}
\gamma_* = (-i)^{m+1} \gamma_0 \gamma_1 \dots \gamma_{D-1} .
\end{equation}
which satisfies
\begin{equation}\label{eq:gammastarprop}
(\gamma_*)^2 = I, \quad \{ \gamma_*, \gamma_\mu \} = 0, \quad (\gamma_*)^\dagger = \gamma_* .
\end{equation}
It makes the link between rank $r$ and rank $D-r$ antisymmetric products of gamma matrices,
\begin{equation} \label{eq:evenduality}
\gamma^{\mu_1 \dots \mu_r} = - \frac{(-i)^{m+1}}{(D-r)!} \varepsilon^{\mu_r \dots \mu_1 \nu_1 \dots \nu_{D-r}} \gamma_{\nu_1 \dots \nu_{D-r}} \gamma_*
\end{equation}
(notice the index ordering).
In odd dimensions $D=2m+1$, there is no $\gamma_*$ and the analogue of this relation is
\begin{equation}
\gamma^{\mu_1 \dots \mu_r} = \frac{i^{m+1}}{(D-r)!} \varepsilon^{\mu_1 \dots \mu_D} \gamma_{\mu_D \dots \mu_{r+1}} .
\end{equation}
We use the convention $\varepsilon_{012\dots (D-1)} = +1 = - \varepsilon^{012\dots (D-1)}$ for the totally antisymmetric $\varepsilon$ tensor. Using these relations, one can prove the identities
\begin{align}
\varepsilon^{ijl_1 \dots l_{d-2}} \gamma_{l_1 \dots l_{d-2}} &= i^{m+1} (d-2)! \gamma^{ij} \gamma_0 \hat{\gamma}, \label{eq:gij} \\
\gamma^0 \gamma^{ijk} &= \frac{(-i)^{m+1}}{(d-3)!} \varepsilon^{ijk l_1 \dots l_{d-3}} \gamma_{l_1 \dots l_{d-3}} \hat{\gamma}, \label{eq:gijk}
\end{align}
where $m = \lfloor D/2 \rfloor$ and where we define $\hat{\gamma}$ to be the chirality matrix $\gamma_*$ in even space-time dimension and the identity matrix in odd space-time dimension,
\begin{equation}
m = \lfloor D/2 \rfloor = \lfloor (d+1)/2 \rfloor, \quad \, \hat{\gamma} = \left\{ \begin{array}{ll}
\gamma_* &\text{ if $D$ is even} \\
I &\text{ if $D$ is odd.}
\end{array} \right.
\end{equation}
The spatial $\varepsilon$ tensor is $\varepsilon_{12\dots d} = +1 = \varepsilon^{12\dots d}$, and spatial indices are contracted with the spatial metric $\delta_{ij}$.
With these definitions, equations \eqref{eq:gij} and \eqref{eq:gijk} are valid in all dimensions. The $\hat{\gamma}$ matrix satisfies
\begin{equation}
(\hat{\gamma})^\dagger = \hat{\gamma}, \quad \hat{\gamma}^2 = I, \quad \hat{\gamma} \gamma_\mu = (-1)^{D+1}\, \gamma_\mu \hat{\gamma} = (-1)^{d}\, \gamma_\mu \hat{\gamma}.
\end{equation}
Using relation \eqref{eq:gij} and its dual, the following identity can also be proved:
\begin{equation}
\varepsilon_{i_1 \dots i_{d-1}j} \gamma^{jk} = \frac{1}{2}(-1)^{d-1}(d-1)\, \delta^k_{[i_1} \varepsilon_{i_2 \dots i_{d-1}] pq} \gamma^{pq} .
\end{equation}
It is the generalization of equation (C.22) of reference \cite{Henneaux:2017xsb} to arbitrary dimension.

\providecommand{\href}[2]{#2}\begingroup\raggedright\endgroup

\end{document}